\documentclass[preprint]{elsarticle}
\title{Lambda-Free Logical Frameworks\tnoteref{support}}
\tnotetext[support]{Corresponding address: Department of Computer Science, Royal Holloway, University of London, Egham Hill, Egham, Surrey.  TW20 0EX.  England.
Tel.: +44 1784 443421.  Fax: +44 1784 439786.
Email \texttt{robin@cs.rhul.ac.uk} \\
This research was supported by the UK EPSRC research grant Pythagoras GR/R84092, the EU Framework VI grant TYPES 510996, and the UK EPSRC research fellowship EP/D066638/1.}
\author[rhul]{Robin Adams}
\ead{robin@cs.rhul.ac.uk}
\address[rhul]{Royal Holloway, University of London}

\usepackage{amssymb}
\usepackage{prooftree}
\usepackage{amsmath}
\usepackage{diagrams}
\usepackage{eqnarray}
\usepackage{QED}

\newcommand{\abs}{\hat{\,}}
\newcommand{\Type}{\mathbf{Type}}
\newcommand{\El}[1]{\mathrm{El} \left( {#1} \right)}
\newcommand{\brackets}[1]{\left[ \! \left[ {#1} \right] \! \right]}
\newcommand{\vald}{\ \mathrm{valid}}
\newcommand{\kind}{\ \mathrm{kind}}
\newcommand{\dom}{\operatorname{dom}}
\newcommand{\lift}{\mathrm{lift}}
\newcommand{\NF}{\mathrm{NF}}
\newcommand{\Ar}{\mathrm{Ar}}
\newcommand{\TFk}{\mathrm{TF}_{\mathrm{k}}}
\newcommand{\lbl}{\mathcal{L}}
\newcommand{\lft}{\mathrm{lift}}
\newcommand{\FV}[1]{\mathrm{FV} \left( {#1} \right)}

\newcommand{\CUT}[3]{\mathrm{CUT} \left( {#1} , {#2}, {#3} \right)}
\newcommand{\FUNC}[3]{\mathrm{FUNC} \left( {#1} , {#2}, {#3} \right)}
\newcommand{\GFUNC}[3]{\mathrm{GFUNC} \left( {#1} , {#2}, {#3} \right)}
\newcommand{\FUNCEQ}[3]{\mathrm{FUNCEQ} \left( {#1} , {#2}, {#3} \right)}
\newcommand{\GFUNCEQ}[3]{\mathrm{GFUNCEQ} \left( {#1} , {#2}, {#3} \right)}
\newcommand{\CC}[3]{\mathrm{CC} \left( {#1} , {#2}, {#3} \right)}
\newcommand{\EQVAL}[2]{\mathrm{EQVAL} \left( {#1} , {#2} \right)}
\newcommand{\LF}{\ensuremath{\mathrm{LF}}}
\newcommand{\boxes}[1]{\left[ \! \left[ {#1} \right] \! \right]}

\newtheorem{thm}{Theorem}[section]
\newtheorem{lm}[thm]{Lemma}

\newtheorem{prop}[thm]{Proposition}
\newdefinition{df}[thm]{Definition}
\newtheorem{conj}[thm]{Conjecture}

\newproof{pf}{Proof}

\begin{document}

\begin{abstract}
We present the definition of the logical framework TF, the \emph{Type Framework}.  TF is a lambda-free logical framework; it does not
include lambda-abstraction or product kinds.  We give formal proofs of several results in the metatheory of TF, and show how it can be conservatively
embedded in the logical framework LF: its judgements can be seen as the judgements of LF that are in beta-normal, eta-long normal form.  We show how several
properties, such as the injectivity of constants and the strong normalisation of an object theory, can be proven more easily in TF, and then `lifted' to LF.
\end{abstract}


\maketitle

\section{Introduction}

A \emph{logical framework} is a typing system intended as a meta-language for the specification of other formal systems, which may themselves be type theories or other systems of logic, such as predicate logic.  Traditionally, logical frameworks are based on a typed lambda calculus; variable binding is represented by lambda-abstraction in the framework, and substitution by application in the framework.  The correspondence between the object theory and its representation in the framework is not exact: each entity of the object theory is represented by more than one object in the framework --- typically, $\beta \eta$-convertible objects represent the same entity of the object theory --- and there are objects in the framework (such as partially applied meta-functions) that do not correspond to any entity of the object theory.  It is therefore necessary to prove \emph{adequacy theorems} establishing the relationship between an object theory and its representation in a logical framework; and these theorems are notoriously often difficult to prove.

It is possible to construct a logical framework that does not employ all the apparatus of the lambda calculus.  We can construct logical frameworks that do not make use of abstraction and substitution, but instead involve only parametrisation and the instantiation of parameters. We shall call these \emph{lambda-free} logical frameworks.  They can be seen as frameworks that only use $\beta$-normal, $\eta$-long normal forms.  Lambda-free frameworks provide a more faithful representation of an object theory --- there is a one-to-one correspondence between the objects of the framework and the terms and types of the object theory.  Because of this, many results including adequacy theorems are easier to prove in a lambda-free framework.

It is often possible to \emph{embed} a lambda-free framework $L$ within a traditional framework $F$; that is, to provide a translation from $L$ into $F$ such that the derivable judgements of $L$ map onto exactly the derivable judgements of $F$ that are in normal form.  $F$ can then be seen as a conservative extension of $L$.  Once this embedding has been established, we can `lift' results from $L$ to $F$; that is, we can prove a result for $L$, and then deduce that the corresponding result holds for $F$ as a corollary.

There is a price to be paid for using a lambda-free framework: the early metatheoretic results are much more difficult to establish, as is the soundness of the embeddings discussed above.  But this is a `one-time' cost; once this price has been paid, it is comparatively easy to prove many results in the lambda-free framework, and then lift them to the traditional frameworks.

\subsection{Background and Outline}

The term `lambda-free logical framework' was first use to describe the framework PAL$^+$ \cite{luo:palplus}, which uses parametrisation and definitions as its basic notions rather than lambda-abstraction.  In PAL$^+$, however, it is possible to form abstractions (using parametric definition) that can then be applied to objects.

We are using the phrase `lambda-free logical framework' in a stricter sense, to describe a framework which does not permit abstractions to be applied to objects, and which therefore contain no framework-level notion of reduction.  We shall use the phrase `traditional framework' throughout this paper to denote a logical framework that is not lambda-free, such as the Edinburgh LF \cite{hhp:lf} or Martin-L\"of's Logical Framework \cite{NPS:book}.
When we represent a formal system $S$ within a logical framework $F$, the system $S$ is referred to as the \emph{object theory}.


The framework TF first appeared in an unpublished note by Aczel \cite{yalf}.  It was developed by myself in my thesis \cite{adams:thesis}. In particular, 
I introduced the set of arities to organise the grammar, and made explicit the
definition of instantiation.

In Section 2, we give the formal definition of TF, and describe how a type theory may be specified in TF.  In Section 3, we begin to prove the metatheoretic properties of TF.  We would like to prove that these properties hold under an arbitrary type theory specification in TF.  However, for most of the properties considered in Section 3, we are at present only able to prove them for two large classes of specifications --- those with no equation declarations, and those which do not involve variables of order 2 or higher\footnote{In Adams \cite{adams:thesis}, the properties in Section 3 were claimed to hold under an arbitrary specification, but a mistake has since been found in the proof.}.  The proofs are given in Section 3, with the more technical proofs given in the Appendix.

In Section 4, we describe a second lambda-free logical framework $\TFk$, which is a Church-typed version of TF; that is, the bound variables are labelled with their kinds.  We define translations between TF and $\TFk$ in Section 4.  It is often very convenient to have these two versions of TF available, and to be able to move between them at will.

In Section 5, we show how TF may be embedded in LF, a Church-typed version of Martin-L\"of's Logical Framework \cite{luo:car}.  We do so by defining a translation from $\TFk$ to LF and from LF to TF, taking advantage of the results of Section 4.  We show how this embedding allows results to be \emph{lifted}; that is, a result may be proven to hold for TF, and the fact that it holds for LF follows as an easy corollary.  We demonstrate this for two results: the injectivity of type constructors, and strong normalisation of an object theory.

In Section \ref{section:related}, we describe two other frameworks that have appeared in the literature which are lambda-free logical frameworks in the stricter sense: the Concurrent Logical Framework (Concurrent LF) \cite{hl:mmlf, lp:brtslf} and DMBEL \cite{plotkin:afltt, pollack:srlf}.  In both of these frameworks, abstractions may be formed, and a constant or variable may be applied to an abstraction, but abstractions may not themselves be applied to objects.

Each of these may be conservatively embedded in TF. That is, we can find a subsystem $S$ of TF such that there exist bijective translations between Concurrent LF and $S$, and such that TF is a conservative extension of $S$.  Likewise, we can find a subsystem $S'$ such that there exist bijective translations between DMBEL and $S'$, and such that TF is a conservative extension of $S'$.  It is possible to find many such subsystems of TF, which all extend one another conservatively; this idea, called a `modular hierarchy of logical frameworks', was described in Adams \cite{adams:modhie} and the formal details given in Adams \cite{adams:thesis}.  We give the details in the case of Concurrent LF and DMBEL in Section \ref{section:related}.

\paragraph{Abbreviation}  Throughout this paper, the phrase `induction hypothesis' shall be abbreviated to `i.h.'.

\section{The Type Framework $\mathrm{TF}$}
\label{section:TF}

We present our first example of a lambda-free framework, the \emph{Type Framework} TF.  The
framework TF includes nothing but what is essential for representing an object theory.  In particular, it contains neither
lambda-abstraction nor local definition; its basic concepts are
parametrisation, the instantiation of parameters, and the declaration
of equations.

\subsection{Grammar}

\subsubsection{Arities}

We begin by introducing the set of \emph{arities}, with which we shall organise the syntax of TF.  

The arities are defined inductively thus:
\begin{quote}
If $\alpha_1$, \ldots, $\alpha_n$ are arities, then $(\alpha_1, \ldots, \alpha_n)$ is an arity.
\end{quote}

The base case of this definition is the case $n=0$, yielding the arity $()$, which we shall write as $\mathbf{0}$.  The next arities that can be formed are
\[ \overbrace{(\mathbf{0}, \ldots, \mathbf{0})}^{n} \]
for positive $n$; we shall write this arity as $\mathbf{n}$.  The next arities that can be formed are $(\mathbf{n_1}, \ldots, \mathbf{n_k})$, and so forth.

The intuition behind the arities is that an $(\alpha_1, \ldots, \alpha_n)$-ary function is a function that takes $n$ arguments --- namely an $\alpha_1$-ary function, \ldots, and an $\alpha_n$-ary function --- and returns an entity (term or type) of the object theory.  In particular, a $\mathbf{0}$-ary (or \emph{base}) function is just an entity of the object theory; a $\mathbf{2}$-ary function is a binary operation on the entities of the object theory; and so forth.

We denote by $\alpha \abs \beta$ the \emph{concatenation} of the two arities $\alpha$ and $\beta$:
\[ (\alpha_1, \ldots, \alpha_m) \abs (\beta_1, \ldots, \beta_n) \equiv (\alpha_1, \ldots, \alpha_m, \beta_1, \ldots, \beta_n) \enspace . \]

We also ascribe an \emph{order} to each arity as follows:
\begin{itemize}
\item
The only 0th-order, or \emph{base}, arity is $\mathbf{0}$.
\item
If the highest order among the arities $\alpha_1$, \ldots, $\alpha_n$ is $k$, then $(\alpha_1, \ldots, \alpha_n)$ is a $k+1$st-order arity.
\end{itemize}
For example, the first-order arities are those of the form $\mathbf{n}$ for positive $n$, and the second-order arities are those of the form $(\mathbf{n_1}, \ldots, \mathbf{n_k})$ where at least one $n_i$ is positive.

We say the arity $\alpha$ is a \emph{subarity} of the arity $\beta$ if $\alpha$ occurs inside $\beta$.  We say $\alpha$ is a \emph{proper} subarity of $\beta$ if $\alpha$ is a subarity of $\beta$ and $\alpha \not\equiv \beta$.

\subsection{Objects}

The \emph{objects} of TF are expressions intended to represent the terms and types of the object theory.  They are built up from \emph{variables} and \emph{constants}, to each of which is assigned an arity.  The constants shall be used for the type constructors and term constructors of the object theory.  The variables shall be used as the variables of the object theory.

\pagebreak

The set of objects is defined by the following inductive definition:
\begin{quote}
If $z$ is an $\alpha$-ary constant or variable, where
\[ \alpha \equiv ((\alpha_{11}, \ldots, \alpha_{1r_1}), \ldots, (\alpha_{n1}, \ldots, \alpha_{nr_n})) \enspace , \]
then
\begin{equation}
\label{eq:object}
 z[[x_{11}, \ldots, x_{1r_1}]M_1, \ldots, [x_{n1}, \ldots, x_{nr_n}]M_n]
\end{equation}
is an object, where each $x_{ij}$ is an $\alpha_{ij}$-ary variable, and each $M_i$ an object.  Each $x_{ij}$ is bound within the corresponding object $M_i$, and we identify objects up to $\alpha$-conversion.
\end{quote}

The base case of this definition is that, if $z$ is a base variable or constant (that is, a $\mathbf{0}$-ary variable or constant), then $z[]$ is an object; we shall henceforth write this object as just $z$.  Likewise, if $z$ is an $\mathbf{n}$-ary variable or constant, then $z[[]M_1, \ldots, []M_n]$ is an object for any objects $M_1$, \ldots, $M_n$; we shall write this object simply as $z[M_1, \ldots, M_n]$.

The subexpressions of the object (\ref{eq:object}) such as $[x_1, \ldots, x_r]M$ are not first-class entities of TF; they cannot occur except as arguments to some variable or constant $z$.  Nevertheless, it shall be convenient to have some way of referring to these pieces of raw syntax.  We shall therefore introduce the following terminology:
\begin{itemize}
 \item
An $(\alpha_1, \ldots, \alpha_n)$-ary \emph{variable sequence} is a sequence of $n$ distinct variables $\langle x_1, \ldots, x_n \rangle$, where $x_i$ has arity $\alpha_i$.
\item
An $\alpha$-ary \emph{abstraction} is an expression of the form $[\vec{x}]M$, where $\vec{x}$ is an $\alpha$-ary variable sequence, and $M$ an object.  We take each member of $\vec{x}$ to be bound within this abstraction, and identify abstractions up to $\alpha$-conversion.
\item
An $(\alpha_1, \ldots, \alpha_n)$-ary \emph{abstraction sequence} is a sequence $\langle F_1, \ldots, F_n \rangle$, where $F_i$ is an $\alpha_i$-ary abstraction.
\end{itemize}
Thus, an object has the form $z[\vec{F}]$, where $z$ is an $\alpha$-ary variable or constant, and $\vec{F}$ an $\alpha$-ary abstraction sequence.  We shall often write this object as just $z \vec{F}$.

We note that the only expressions that can occur as arguments to a symbol are abstractions.  In the situations where we would naturally wish to write a variable or constant in an argument position, we instead write its \emph{$\eta$-long form}.

\begin{df}[$\eta$-long Form]
Given any $\alpha$-ary variable or constant $z$, the \emph{$\eta$-long form} $z^\eta$ of $z$ is the $\alpha$-ary abstraction defined by recursion on $\alpha$ as follows:

If $\alpha \equiv (\alpha_1, \ldots, \alpha_n)$, then
\[ z^\eta \equiv [x_1, \ldots, x_n] z[x_1^\eta, \ldots, x_n^\eta] \enspace , \]
where each $x_i$ is an $\alpha_i$-ary variable.  (By $\alpha$-conversion, it does not matter which variables we choose.)
\end{df}

\subsection{Hereditary Substitution and Employment}

We cannot use the familiar operation of substitution in TF.  The result of substituting an abstraction $[\vec{y}]M$ for the variable $x$ in the object $x \vec{F}$ is not an object of TF; rather, it would be a $\beta$-redex.

Instead, we introduce an operation that we name \emph{instantiation}.  The operation of \emph{instantiating} an abstraction $F$ for a variable $x$ can be thought of as substituting $F$ for $x$, then \emph{reducing to normal form} (that is, $\beta$-normal, \linebreak $\eta$-long form).  However, we note that the definition does not use any notion of reduction.

\begin{df}[Instantiation]
 Given an $\alpha$-ary abstraction $F$, an $\alpha$-ary variable $x$, and an object $N$, the object $\{ F / x \} N$, the result of \emph{instantiating} $F$ for $x$ in $N$, is defined by recursion firstly on the arity $\alpha$, secondly on the object $N$, as follows:
\[  \{ F / x \} z [G_1, \ldots, G_n] \equiv z[\{F/x\}G_1, \ldots, \{F/x\}G_n] \qquad (z \not\equiv x) \]
If $F \equiv [t_1, \ldots, t_n]P$, then
\[ \{ F / x \} x [G_1, \ldots, G_n] \equiv
\{ \{ F/x \} G_1 / t_1 \} \cdots \{ \{ F/x \} G_n / t_n \} P \enspace . \]
We assume here, through $\alpha$-conversion, that no $t_i$ occurs free in any $G_j$.
\end{df}

We shall also introduce a notational convention that shall play the role of abstraction: if $x$ is an $\alpha$-ary variable and $F$ a $\beta$-ary abstraction, then $[x]F$ is an $(\alpha) \abs \beta$-ary abstraction, defined by
\[ [x][y_1, \ldots, y_n]M \equiv [x,y_1,\ldots,y_n]M \enspace . \]

Finally, we define an operation, which we shall call \emph{employment}, to play the role usually taken by application.  The result of \emph{employing} $F$ on $G$, denoted $F \bullet G$, can be thought of as the normal form of the application $FG$.  The definition is:

\begin{df}[Employment]
Given an $(\alpha) \abs \beta$-ary abstraction $[x]F$ and an $\alpha$-ary abstraction $G$, the $\beta$-ary abstraction $F \bullet G$, the result of \emph{employing} $[x]F$ on $G$, is defined by
\[ ([x]F) \bullet G \equiv \{G/x\}F \enspace . \]
\end{df}
We have used our newly introduced notation $[x]M$ in this definition; written out in full, the above equation is
\[ ([x,y_1, \ldots, y_n]M) \bullet G \equiv [y_1, \ldots, y_n]\{G/x\}M \enspace . \]

We shall abbreviate the repeated use of employment as follows: if $\vec{G}$ is the abstraction sequence $\langle G_1, \ldots, G_n \rangle$, then $F \bullet \vec{G}$ abbreviates $F \bullet G_1 \bullet G_2 \bullet \cdots \bullet G_n$, that is,
\[ (( \cdots (F \bullet G_1) \bullet G_2) \bullet \cdots ) \bullet G_n \enspace . \]

\paragraph{Remark}
We note that there is a strong correspondence between our syntax and the simply-typed lambda calculus.
Our arities correspond to the types of the simply-typed lambda calculus, and our abstractions to the terms.  Instantiation corresponds to the strategy of innermost reduction.  Thus, the fact that our definition of instantiation is total corresponds to the fact that the simply-typed lambda calculus is weakly normalisable.

\subsection{Kinds}

A \emph{base kind} in TF is either the symbol $\Type$, or has the form $\El{A}$ for some object $A$.  The intention is that each type $T$ of the object theory is represented by an object $\brackets{T}$ of kind $\Type$; the terms of type $T$ are then represented by the objects of kind $\El{\brackets{T}}$.

In addition to these, we introduce a set of \emph{$\alpha$-ary product kinds} for every arity $\alpha$.  These shall be used to give kinds to the variables and constants of higher arity.  The definition is by recursion on $\alpha$:
\begin{quote}
 An $(\alpha_1, \ldots, \alpha_n)$-ary product kind is an expression of the form
\begin{equation}
 \label{eq:prodkind}
(x_1 : K_1, \ldots, x_n : K_n) T
\end{equation}
where the $x_i$s are distinct variables, $x_i$ being of arity $\alpha_i$; each $K_i$ is an $\alpha_i$-ary product kind; and $T$ is a base kind.
\end{quote}
We take each variable $x_i$ to be bound within $K_{i+1}$, $K_{i+2}$, \ldots, $K_n$ and $T$ in this product kind, and identify product kinds up to $\alpha$-conversion.

The intuition is that the kind (\ref{eq:prodkind}) represents the collection of functions that take $n$ arguments --- namely $F_1$ of kind $K_1$, $F_2$ of kind $\{F_1/x_1\}K_2$, \ldots, and $F_n$ of kind $\{F_1/x_1, \ldots, F_{n-1}/x_{n-1}\}K_n$ --- and returns an object of kind $\{F_1/x_1, \ldots, F_n/x_n\}T$.

If $K \equiv (x_1 : K_1, \ldots, x_n : K_n) T$, then we shall write $(y : J)K$ for \linebreak $(y : J, x_1 : K_1, \ldots, x_n : K_n) T$.

Just as with abstractions, so the product kinds of non-zero arity are not considered first-class entities of TF; only the base kinds are.  We shall however make use of the higher product kinds  to give kinds to the variables and constants of higher arity.  We shall even talk of an abstraction being a member of a product kind; however, this shall not be represented by a primitive judgement form of TF.

\paragraph{Contexts}
A \emph{context} $\Gamma$ in TF is a sequence of the form:
\[ x_1 : K_1, \ldots, x_n : K_n \]
where the $x_i$s are distinct variables, and each $x_i$ has the same arity as the corresponding product kind $K_i$.  If each $x_i$ has arity $\alpha_i$, we say the context $\Gamma$ has arity $(\alpha_1, \ldots, \alpha_n)$, and its order $o(\Gamma)$ is then the order of $(\alpha_1, \ldots, \alpha_n)$.   The variable sequence $\langle x_1, \ldots, x_n \rangle$ is called the \emph{domain} of the context $\Gamma$, $\dom \Gamma$.

Thus, an $\alpha$-ary kind has the form $(\Gamma)T$, where $\Gamma$ is an $\alpha$-ary context and $T$ a base kind.

\subsection{Judgement Forms}

There are three \emph{primitive judgement forms} in TF:
\begin{eqnarray*}
\lefteqn{\Gamma \vald} \\
\Gamma & \vdash & M : T \\
\Gamma & \vdash & M = N : T
\addtocounter{equation}{1}\end{eqnarray*}
where $\Gamma$ is a context, $M$ and $N$ are objects, and $T$ is a base kind.  These are intended to express that $\Gamma$ is a valid context; that the object $M$ has kind $T$ under the context $\Gamma$; and that the objects $M$ and $N$ are equal objects of kind $T$ under $\Gamma$, respectively.

We now introduce \emph{defined judgement forms} to deal with the abstractions and product kinds of higher arity:
\[ \Gamma \Vdash K \kind; \qquad \Gamma \Vdash K = K'; \qquad \Gamma \Vdash F : K; \qquad \Gamma \Vdash F = G : K \enspace . \]
Each of these judgements is defined to be a set of primitive judgements.  We shall always use the double turnstile $\Vdash$ to indicate a defined judgement form.

For any base kind $T$, the defined judgement $\Gamma \Vdash T \kind$ is defined as follows:
\begin{eqnarray*}
(\Gamma \Vdash \Type \kind) & = & \{ \Gamma \vald \}  \\
(\Gamma \Vdash \El{A} \kind) & = & \{ \Gamma \vdash A : \Type \}
\addtocounter{equation}{1}\end{eqnarray*}
For any $\alpha$-ary product kind $K$, the judgement $\Gamma \Vdash K \kind$ is defined by:
\[ (\Gamma \Vdash (\Delta)T \kind) = (\Gamma, \Delta \Vdash T \kind) \enspace . \]
Equality of base kinds is defined by:
\begin{eqnarray*}
(\Gamma \Vdash \Type = \Type) & = & \{ \Gamma \vald \}\\
(\Gamma \Vdash \El{A} = \El{B}) & = & \{ \Gamma \vdash A = B : \Type \}
\addtocounter{equation}{1}\end{eqnarray*}
We leave `$\Gamma \Vdash \Type = \El{B}$' and `$\Gamma \Vdash \El{A} = \Type$' undefined.

Equality of product kinds and contexts is defined recursively by
\begin{eqnarray*}
(\Gamma \Vdash (\Delta) T = (\Delta') T') & = & (\Gamma \Vdash \Delta = \Delta') \cup \{ \Gamma \Vdash T = T' \} \\
& &  \\
(\Gamma \Vdash \langle \rangle = \langle \rangle) & = & \{ \Gamma \vald \} \\
(\Gamma \Vdash \Delta, x : K = \Delta', x : K') & = & (\Gamma \Vdash \Delta = \Delta') \cup (\Gamma, \Delta \Vdash K = K')
\addtocounter{equation}{1}\end{eqnarray*}

For example, the defined judgement $\Gamma \Vdash (x:A)B = (x:C)D$ is defined to be the set
\[ \{ \Gamma \vald, \; \Gamma \vdash A = C : \Type, \; \Gamma, x:A \vdash B = D : \Type \} \enspace . \]
 The judgement $\Gamma \Vdash (x:A)B = (x:C) \Type$ is undefined.

We introduce defined judgement forms $\Gamma \Vdash F : K$ and $\Gamma \Vdash F = G : K$ for the inhabitation of a product kind $K$ by an abstraction $F$, and the equality of two abstractions $F$ and $G$ of product kind $K$; here, $F$, $G$ and $K$ must all have the same arity.
\begin{eqnarray*}
(\Gamma \Vdash [\vec{x}]M : (\Delta) T) & = & \{ \Gamma, \Delta \vdash M : T \} \\
(\Gamma \Vdash [\vec{x}]M = [\vec{x}]N : (\Delta)T) & = & \{ \Gamma, \Delta \vdash M = N : T \}
\addtocounter{equation}{1}\end{eqnarray*}
We assume here that we have applied $\alpha$-conversion to ensure that the same variable sequence $\vec{x}$ is used in both $[\vec{x}]P$ and $[\vec{x}]Q$, and is also the domain of the context $\Delta$.

Finally, we introduce judgement forms
\begin{itemize}
 \item 
 $\Gamma \Vdash \vec{F} :: \Delta$, denoting that $\vec{F}$ \emph{satisfies} the context $\Delta$; that is, $\vec{F}$ is a sequence of abstractions whose kinds are those given by the context $\Delta$;
\item $\Gamma \Vdash \vec{F} = \vec{G} :: \Delta$, denoting that $\vec{F}$ and $\vec{G}$ are two equal abstraction sequences that satisfy $\Delta$.
\end{itemize}
The judgement forms are defined as follows:
\begin{eqnarray*}
(\Gamma \Vdash \langle \rangle :: \langle \rangle) & = & \{ \Gamma \vald \} \\
(\Gamma \Vdash \vec{F}, F_0 :: \Delta, x : K) & = & (\Gamma \Vdash \vec{F} :: \Delta) \cup ( \Gamma \Vdash F_0 : \{ \vec{F} / \Delta \} K ) \\
& &  \\
\Gamma \Vdash \langle \rangle = \langle \rangle :: \langle \rangle & = & \{ \Gamma \vald \} \\
\lefteqn{\Gamma \Vdash (\vec{F}, F_0) = (\vec{G}, G_0) :: (\Delta, x : K)} \\
 & = & (\Gamma \Vdash \vec{F} = \vec{G} :: \Delta) \cup ( \Gamma \Vdash F_0 = G_0 : \{ \vec{F} / \Delta \} K )
\addtocounter{equation}{1}\end{eqnarray*}

\subsection{Rules of Deduction}

We are finally able to give the rules of deduction of TF.  They are listed in Figure \ref{fig:TF}.  They consist of the rules (emp) and (ctxt) determining when a context is valid; (var) and (var\_eq), the typing and congruence rules for the application of a variable; (ref), (sym) and (trans), which ensure that the judgemental equality is an equivalence relation; and (conv) and (conv\_eq), which ensure that equal kinds have the same objects.

We note in passing how few rules there are compared to logical frameworks of similar expressiveness such as LF \cite{luo:car} and ELF \cite{hhp:lf}.  In particular, the two rules (var) and (var\_eq) do all the work normally done by the rules governing typing and congruence of applications and abstractions, and $\beta$- and $\eta$-contractions.  We have shifted this burden from the rules of deduction to the syntax.

\begin{figure}
\[ \begin{array}{rcl}
(\mbox{emp\_ctxt}) & \begin{prooftree}
    \justifies
\langle \rangle \vald
   \end{prooftree} & \\
& & \\
(\mbox{ctxt}) & \begin{prooftree}
 \Gamma \Vdash K \kind
\justifies
\Gamma, x : K \vald
\end{prooftree} & (x \notin \dom \Gamma) \\
& & \\
(\mbox{var}) & \begin{prooftree}
\Gamma \Vdash \vec{F} :: \Delta
\justifies
\Gamma \vdash x \vec{F} : \{ \vec{F} / \Delta \} T
   \end{prooftree} &
(x : (\Delta)T \in \Gamma) \\
& & \\
(\mbox{var\_eq}) & \begin{prooftree}
 \Gamma \Vdash \vec{F} = \vec{G} :: \Delta
\justifies
\Gamma \vdash x \vec{F} = x \vec{G} : \{ \vec{F} / \Delta \} T
\end{prooftree} &
(x : (\Delta) T \in \Gamma) \\
& & \\
(\mbox{ref}) & \begin{prooftree}
\Gamma \vdash M : T
\justifies
\Gamma \vdash M = M : T
\end{prooftree} & \\
& & \\
(\mbox{sym}) & \begin{prooftree}
 \Gamma \vdash M = N : T
\justifies
\Gamma \vdash N = M : T
\end{prooftree} & \\
& & \\
(\mbox{trans}) & \begin{prooftree}
 \Gamma \vdash M = N : T
\quad
\Gamma \vdash N = P : T
\justifies
\Gamma \vdash M = P : T
\end{prooftree} & \\
& & \\
(\mbox{conv}) & \begin{prooftree}
    \Gamma \vdash M : \El{A}
\quad
\Gamma \vdash A = B : \Type
\justifies
\Gamma \vdash M : \El{B}
   \end{prooftree} & \\
& & \\
(\mbox{conv\_eq}) & \begin{prooftree}
 \Gamma \vdash M = N : \El{A}
\quad
\Gamma \vdash A = B : \Type
\justifies
\Gamma \vdash M = N : \El{B}
\end{prooftree} & 
\end{array} \]
\caption{Rules of Deduction of TF}
\label{fig:TF}
\end{figure}


\subsubsection{Type Theory Specifications}
\label{declarations}

An object theory is represented in TF by extending the logical framework with several new rules of deduction, representing the formation of the terms and types of the object theory and the computation rules of the object theory.
\pagebreak

Formally, a \emph{type theory specification} in TF is a set of \emph{declarations}, of two possible forms:
\begin{itemize}
\item
\emph{constant declarations} of the form
\[ c : K \]
where $c$ is a constant and $K$ a kind of the same arity;
and 
\item
\emph{equation declarations} of the form
\[ (\Delta)(M = N : T) \]
where $\Delta$ is a context, $M$ and $N$ objects and $T$ a base kind.
\end{itemize}

The intention is that the constant declarations represent the term- and type-constructors of the object theory, and the equation declarations represent the computation rules of the object theory.

Making the constant declaration $c : (\Delta) T$ has the effect of adding the following two rules of deduction to the framework (c.f.~the rules (var) and (var\_eq)):
\begin{equation*}
\label{eq:constdec}
(\mbox{const}) \; \begin{prooftree}
\Gamma \Vdash \vec{F} :: \Delta
\justifies
\Gamma \vdash c \vec{F} : \{\vec{F} / \Delta\}T
\end{prooftree}
\qquad
(\mbox{const\_eq}) \; \begin{prooftree}
\Gamma \Vdash \vec{F} = \vec{G} :: \Delta
\justifies
\Gamma \vdash c \vec{F} = c \vec{G} : \{ \vec{F} / \Delta \} T
\end{prooftree}
\end{equation*}

Making the equation declaration $(\Delta)(M = N : T)$ has the effect of adding the following rule to the framework:
\begin{equation*}
\label{eq:eqdec}
(\mbox{eq}) \;  \begin{prooftree}
\Gamma \Vdash \vec{F} :: \Delta
\justifies
\Gamma \vdash \{ \vec{F} / \Delta \} M = \{ \vec{F} / \Delta \} N : \{ \vec{F} / \Delta \} T
\end{prooftree}
\end{equation*}

We define the \emph{order} $o(\delta)$ of a declaration as follows: the order of $c : K$ is the order of $K$, and the order of $(\Delta)(M = N : T)$ is the order of $\Delta$.
The \emph{order} $o(\mathcal{T})$ of a type theory specification $\mathcal{T}$ is the largest $n$ such that $\mathcal{T}$ contains a declaration of order $n$, or $\omega$ if there is no such maximum.

\subsection{Representing Object Theories in TF}
TF is intended for representing \emph{type theories} that have judgements of the following forms:
\begin{eqnarray}\addtocounter{equation}{1}
 x_1 : A_1, \ldots, x_n : A_n & \vdash & M : B \label{jf1} \\
x_1 : A_1, \ldots, x_n : A_n & \vdash & M = N : B \label{jf2}
\end{eqnarray}
Given such a type theory $T$ that we wish to represent in TF, we begin by forming the appropriate specification.  There will be one constant declaration for each constructor in the grammar of $T$, and one equation declaration for each computation rule in $T$.
\pagebreak

We make these declarations in such a way that:
\begin{itemize}
 \item the objects of kind $\Type$ correspond to the types of $T$;
\item if the object $M : \Type$ corresponds to the type $A$, then the objects of kind $\El{M}$ correspond to the terms of type $A$;
\item the judgements of $T$ of the form (\ref{jf1}) correspond to the TF judgements of the form
\begin{equation}
 x_1 : \El{A_1}, \ldots, x_n : \El{A_n} \vdash M : \El{B} \enspace ; \label{jf3}
\end{equation}
\item the judgements of $T$ of the form (\ref{jf2}) correspond to the TF judgements of the form
\begin{equation}
 x_1 : \El{A_1}, \ldots, x_n : \El{A_n} \vdash M = N : \El{B} \enspace . \label{jf4}
\end{equation}
\end{itemize}

To specify type theories such as the Calculus of Constructions \cite{ch:coc}, ECC \cite{luo:car} or Martin-L\"of's Type Theory without W-types \cite{NPS:book} requires a second-order specification.  To specify Martin-L\"of's Type Theory with W-types requires a third-order specification.  To specify UTT \cite{luo:car} requires a specification of order $\omega$.  These examples are described in more detail in Adams \cite{adams:thesis}.

Note that the judgements of TF that represent the judgements of the object theory, those of form (\ref{jf3}) or (\ref{jf4}), have first-order contexts.  This will be important in the following section.  For many of the metatheoretic properties we investigate, we shall be able to prove that they hold for judgements with first-order contexts, but they have not yet been proved to hold for judgements with contexts of order $\geq 2$.

\section{Metatheory}
\label{section:metatheory}
We can now begin to investigate the metatheoretical properties of this system.  Many of these properties are more difficult to prove than the corresponding properties of a traditional logical framework; in particular, it is often the case that several properties need to be established simultaneously by a single induction.  This should be seen as the `one-time' cost of using a lambda-free framework.

\subsection{Grammar}

We begin by demonstrating some properties of the operations of instantiation and employment.  Many of them are analogous to properties of substitution in more familiar languages; we shall point out these analogies as we proceed.

\pagebreak

\begin{lm}
Let $\FV{X}$ denote the set of free variables in the object or abstraction $X$.
 \begin{enumerate}
  \item $\FV{\{F/x\}N} \subseteq (\FV{N} \setminus \{x\}) \cup \FV{F}$
\item $\FV{F \bullet G} \subseteq \FV{F} \cup \FV{G}$.
 \end{enumerate}
\end{lm}

\begin{pf}
 Part 1 is proved by induction on the object $N$.  Part 2 follows directly.
\end{pf}

The following is the analogue of the result that, if $x$ is not free in $N$, then $[M/x]N \equiv N$.
\begin{lm}
\label{lm:trivinst}
If $x$ does not occur free in $M$, then
\[ \{F/x\}M \equiv M \enspace . \]
\end{lm}

\begin{pf}
This is easily proven by induction on the object $M$.
\end{pf}

Part 1 of the next lemma is the analogue of the famous Substitution Lemma.
\begin{lm}
Let $\alpha$, $\beta$ and $\gamma$ be arities.  Let $F$ be an $\alpha$-ary abstraction, $G$ a $\beta$-ary abstraction, and $H$ a $(\beta)\abs \gamma$-ary abstraction.  Let $x$ be an $\alpha$-ary variable and $y$ a $\beta$-ary variable, with $x \not\equiv y$.  Let $M$ be an object.
\begin{enumerate}
\item
If $x$ and $y$ are distinct variables, and $y$ does not occur free in $M$, then
\[ \{F/x\}\{G/y\}M \equiv \{\{F/x\}G/y\}\{F/x\}M \enspace . \]
\item
$ \{ F/x \} (H \bullet G) \equiv (\{F/x\}H) \bullet \{F/x\}G$.
\end{enumerate}
\end{lm}

\begin{pf}
Both parts are proved simultaneously by induction on the sum of the orders of $\alpha$ and $\beta$.
\end{pf}

Part 1 of the next lemma is the analogue of the fact that $[M/x]x \equiv M$.  Part 3 is the analogue of the fact that $[x/x]M \equiv M$.
\begin{lm}
\label{lm:eta}
Let $\alpha$ be an arity.
\begin{enumerate}
\item
For any $\alpha$-ary variable $x$ and $\alpha$-ary abstraction $F$,
$\{F/x\} x^\eta \equiv F$.
\item
For any $\alpha$-ary variable $x$ and $\alpha$-ary abstraction sequence $\vec{F}$,
$x^\eta \bullet \vec{F} \equiv x \vec{F}$.
\item
For any $\alpha$-ary variable $x$ and object $M$,
$\{x^\eta / x\}M \equiv M$.
\end{enumerate}
\end{lm}

\begin{pf}
The three parts are proven simultaneously by induction on $\alpha$.  Part 3 requires a secondary induction on the object $M$.
\end{pf}

\subsection{Metatheoretic Properties}

The following results are true in TF.

\begin{thm}$ $
\begin{enumerate}
 \item 
(\textbf{Context Validity})
Every derivation of a judgement of the form $\Gamma, \Delta \vdash J$ has a subderivation of $\Gamma \vald$.
\item

Every derivation of $\Gamma, x : K, \Delta \vdash J$ has a subderivation of $\Gamma \Vdash K \kind$.
\item

If $\Gamma \vdash J$ is derivable, then every free variable in the judgement body $J$ is in the domain of $\Gamma$.
\item

If $\Gamma, x : K, \Delta \vald$, then every free variable in $K$ is in the domain of $\Gamma$.
\item
(\textbf{Weakening})
If $\Gamma \vdash J$, $\Gamma \subseteq \Delta$ and $\Delta \vald$, then $\Delta \vdash J$.
\item
(\textbf{Generation})
If $\Gamma \vdash x \vec{F} : T$, then there is a declaration $x : (\Delta) S$ in $\Gamma$, where
\[ \Gamma \Vdash \vec{F} :: \Delta, \qquad \Gamma \Vdash \{ \vec{F} / \Delta \} S = T \enspace . \]
\item
(\textbf{Generation})
If $\Gamma \vdash c \vec{F} : T$, then a constant declaration $c : (\Delta) S$ has been made, where
\[ \Gamma \Vdash \vec{F} :: \Delta, \qquad \Gamma \Vdash \{ \vec{F} / \Delta \} S = T \enspace . \]
\item
If $\Gamma \vdash M : T$ and $\Gamma \vdash M : T'$, then $\Gamma \Vdash T = T'$.
\end{enumerate}
\end{thm}

\begin{pf}
The first 7 parts are each proved by a simple induction on derivations.  Part 8 follows easily from parts 6 and 7.
\end{pf}

The other metatheoretic properties of TF are very difficult to establish.  We have not been able to prove the following properties in full generality, but only under a set of restrictions on the type theory specification and context.

\begin{df}[Good Specification]
 Let $\mathcal{T}$ be a type theory specification in TF.
\begin{enumerate}
\item We say that $\mathcal{T}$ is \emph{orderable} iff there exists a well-ordering $\prec$ on the declarations of $\mathcal{T}$ such that:
\begin{enumerate}
\item
For every constant declaration $\delta \equiv (c : (\Delta) T)$, it is possible to derive $\Delta \Vdash T \kind$ using only the declarations $\delta'$ such that $\delta' \prec \delta$.
\item For every equation declaration $\delta \equiv (\Delta)(M = N : T)$, it is possible to derive $\Delta \vdash M : T$, $\Delta \vdash N : T$ and $\Delta \Vdash T \kind$ using only the declarations $\delta'$ such that $\delta' \prec \delta$.
\end{enumerate}
 \item We say that $\mathcal{T}$ is \emph{$n$-good} iff, whenever $\Gamma$ is a context of order $\leq n$ and $\Gamma \vdash M = N :T$, then $\Gamma \vdash M : T$ and $\Gamma \vdash N : T$.
\item We say that $\mathcal{T}$ is \emph{good} iff $\mathcal{T}$ is $n$-good for every natural number $n$.
\end{enumerate}
\end{df}

\pagebreak

It is difficult to find general conditions under which we can prove that a specification is good.  So far, we are able to do so for two large classes of specifications:
\begin{thm}$ $
 \begin{enumerate}
  \item If $\mathcal{T}$ contains no equation declarations, then $\mathcal{T}$ is good.
\item If $\mathcal{T}$ is orderable and $o(\mathcal{T}) \leq 2$, then $\mathcal{T}$ is 2-good.
 \end{enumerate}
\end{thm}

\begin{pf}$ $
\begin{enumerate}
 \item A simple proof by induction on derivations shows that, whenever $\Gamma \vdash M = N : T$, then $M \equiv N$ and $\Gamma \vdash M : T$.
\item See Appendix \ref{appendix:2good}.
\end{enumerate}
\end{pf}

\begin{thm}
\label{thm:mainresult}
Let $\mathcal{T}$ be a type theory specification.  Suppose $\mathcal{T}$ is $n$-good, and $\Gamma, x : K, \Delta$ is a context of order $\leq n$.
\begin{enumerate}
 \item
(\textbf{Cut})
If $\Gamma, x : K, \Delta \vdash J$ and $\Gamma \Vdash F : K$ then $\Gamma, \{F/x\} \Delta \vdash \{F/x\}J$.
\item
(\textbf{Functionality})
If $\Gamma, x : K, \Delta \vdash M : T$ and $\Gamma \Vdash F = G : K$ then $\Gamma, \{ F / x \} \Delta \vdash \{F/x\}M = \{G/x\}M : \{F/x\}T$.
\item
(\textbf{Context Conversion})
If $\Gamma, x : K, \Delta \vdash J$ and $\Gamma \Vdash K = K'$ then $\Gamma, x : K', \Delta \vdash J$.
\end{enumerate}
\end{thm}

\begin{pf}
 See Appendix \ref{appendix:A}.
\end{pf}

Once we have got past this hurdle, other properties of TF follow rapidly.

\begin{thm}[Type Validity]
 Suppose that $\mathcal{T}$ is an $n$-good specification, and
\begin{itemize}
 \item 
 for every constant declaration $c : K$ in $\mathcal{T}$, we have $\Vdash K \kind$;
\item
for every equation declaration $(\Delta)(M = N : T)$ in $\mathcal{T}$, we have $\Delta \Vdash T \kind$.
\end{itemize}
Then, whenever $o(\Gamma) \leq n$, if $\Gamma \vdash M : \El{A}$ or $\Gamma \vdash M = N : \El{A}$, we have $\Gamma \vdash A : \Type$.
\end{thm}

\begin{pf}
The proof is by induction on derivations.  The cases (const) and \linebreak (const\_eq) use the first hypothesis with Cut and Functionality respectively.  The case (eq) uses the second hypothesis with Cut.  The other cases are all trivial.
\end{pf}

\begin{thm}[Kind Validity]
Suppose $\mathcal{T}$ is $n$-good and $o(\Gamma) \leq n$.
Then the  following rules are admissible.
\[ \begin{prooftree}
\Gamma \Vdash F : K
\justifies
\Gamma \Vdash K \kind
\end{prooftree}
\qquad
\begin{prooftree}
\Gamma \Vdash F = G : K
\justifies
\Gamma \Vdash K \kind
\end{prooftree} \]
\end{thm}

\begin{pf}
Both rules are proved admissible simultaneously by induction on the derivation of the premise.  The case of the rule (var\_eq) requires Equation Validity.
\end{pf}

\section{The Church-Typed $\mathrm{TF}$}
\label{section:TFk}

The version of TF we have described is \emph{Curry-typed}; that is, the bound variables in abstractions are not annotated with their kinds.  We can also construct a \emph{Church-typed} version of TF, in which objects have the form
\[ z[[x_{11} : K_{11}, \ldots, x_{1r_1} : K_{1r_1}] M_1, \ldots, [x_{n1} : K_{n1}, \ldots, x_{nr_n} : K_{nr_n}]M_n] \enspace . \]
We shall call the Church-typed version of TF by the name $\TFk$\footnote{The `k' here stands for `kind', as we include the kind labels in abstractions.  This system was named $\mathrm{TF}_{\mathrm{c}}$ in Adams \cite{adams:thesis}, the `c' standing for `Church'.  I have decided to abandon this name, as `c' could just as well stand for `Curry'!}.  In this section, we shall give the definition of $\TFk$, prove its metatheoretic properties, and define mutually inverse translations between TF and $\TFk$ that show that the two systems are in some sense equivalent.

It is very convenient to have available two versions of a lambda-free logical framework, and to be able to switch between them at will.  For example, when embedding a lambda-free framework in a traditional framework, it is easier to define translations \emph{into} the Curry-typed version, and \emph{from} the Church-typed version.  We shall be in just this situation when we come to embed TF in LF.

%

\subsection{Grammar}

In $\TFk$, the sets of \emph{objects}, \emph{abstractions}, \emph{abstraction sequences}, \emph{contexts} and \emph{kinds} are all defined simultaneously as follows.
\begin{description}
\item[Objects]
An \emph{object} has the form $z \vec{F}$, where $z$ is an $\alpha$-ary variable or constant and $\vec{F}$ an $\alpha$-ary abstraction sequence, for some arity $\alpha$.
\item[Abstractions]
An \emph{$\alpha$-ary abstraction} has the form $[\Delta]M$, where $\Delta$ is an $\alpha$-ary context and $M$ an object.
\item[Abstraction Sequences]
An \emph{$(\alpha_1, \ldots, \alpha_n)$-ary abstraction sequence} has the \linebreak form $\langle F_1, \ldots, F_n \rangle$, where each $F_i$ is an $\alpha_i$-ary abstraction.
\item[Contexts]
An \emph{$(\alpha_1, \ldots, \alpha_n)$-ary context} has the form $x_1 : K_1, \ldots, x_n : K_n$, where each $x_i$ is an $\alpha_i$-ary variable and $K_i$ an $\alpha_i$-ary kind, with the $x_i$s all distinct.
\item[Kinds]
An \emph{$\alpha$-ary kind} has the form $(\Delta)\Type$ or $(\Delta)\El{M}$, where $\Delta$ is an $\alpha$-ary context and $M$ an object.
\end{description}
In an abstraction $[x_1 : K_1, \ldots, x_n : K_n]M$ or a kind $(x_1 : K_1, \ldots, x_n : K_n)T$, each variable $x_i$ is bound wherever it occurs in $K_{i+1}$, $K_{i+2}$, \ldots, $K_n$, and $M$.  We identify all these expressions up to $\alpha$-conversion.

The \emph{$\eta$-long form} of a symbol in $\TFk$ must be defined with reference to some kind.  For $z$ an $\alpha$-ary variable or constant and $K$ an $\alpha$-ary kind, we define the $\alpha$-ary abstraction $z^K$, the $\eta$-long form of $z$ considered as being of kind $K$, by recursion on $\alpha$ as follows.
\[ z^{(x_1 : K_1, \ldots, x_n : K_n)T} \equiv [x_1 : K_1, \ldots, x_n : K_n] z[x_1^{K_1}, \ldots, x_n^{K_n}] \enspace . \]

The definitions of instantiation and employment in $\TFk$ are very similar to those in TF.
\[ \{ F/x\} z[G_1, \ldots, G_n] \equiv z[\{F/x\}G_1, \ldots, \{F/x\}G_n] \qquad (z \not\equiv x) \]
If $F \equiv [t_1 : K_1, \ldots, t_n : K_n]M$,
\begin{eqnarray*}
 \{F/x\}x[G_1, \ldots, G_n] & \equiv & \{ \{F/x\}G_1 / t_1 \} \cdots \{ \{F/x\}G_n / t_n \} M \\
([x : K]F) \bullet G & \equiv & \{G/x\}F
\addtocounter{equation}{1}\end{eqnarray*}

As in TF, there are three primitive judgement forms in $\TFk$:
\[ \Gamma \vald \qquad \Gamma \vdash M : T \qquad \Gamma \vdash M = N : T \]
where $\Gamma$ is a context, $M$ and $N$ objects and $T$ a base kind.

We define the judgement forms $\Gamma \Vdash K \kind$, $\Gamma \Vdash K = K'$ and $\Gamma \Vdash \Delta = \Delta'$ just as we did for TF.

The judgement form $\Gamma \Vdash F : K$, where $F$ is an $\alpha$-ary abstraction and $K$ an $\alpha$-ary kind, is defined as follows.
\[ (\Gamma \Vdash [\Delta]M : (\Delta')T) = (\Gamma \Vdash \Delta' = \Delta) \cup \{ \Gamma, \Delta' \vdash M : T \} \enspace . \]
The judgement form $\Gamma \Vdash F = G : K$, where $F$ and $G$ are $\alpha$-ary abstractions and $K$ an $\alpha$-ary kind, is defined as follows.
\begin{eqnarray*}
\lefteqn{(\Gamma \Vdash [\Delta_1]M = [\Delta_2]N : (\Delta_3)T)} \\
 & = & (\Gamma \Vdash \Delta_3 = \Delta_1) 
 \cup (\Gamma \Vdash \Delta_3 = \Delta_2) 
 \cup \{ \Gamma, \Delta_3 \vdash M = N : T \}
\addtocounter{equation}{1}\end{eqnarray*}
The judgement form $\Gamma \Vdash \vec{F} :: \Delta$, where $\vec{F}$ is an $\alpha$-ary abstraction sequence and $\Delta$ an $\alpha$-ary context, is defined by recursion on $\alpha$ as follows.
\begin{eqnarray*}
(\Gamma \Vdash \langle \rangle :: \langle \rangle) & = & \{ \Gamma \vald \} \\
(\Gamma \Vdash \Vec{F}, F_0 :: \Delta, x : K) & = & (\Gamma \Vdash \vec{F} :: \Delta) \\
& & \cup (\Gamma \Vdash F_0 : \{\vec{F} / \Delta\}K)
\addtocounter{equation}{1}\end{eqnarray*}
The judgement form $\Gamma \Vdash \vec{F} = \vec{G} :: \Delta$, where $\vec{F}$ and $\vec{G}$ are $\alpha$-ary abstraction sequences and $\Delta$ an $\alpha$-ary context, is defined by recursion on $\alpha$ as follows.
\begin{eqnarray*}
(\Gamma \Vdash \langle \rangle = \langle \rangle :: \langle \rangle) & = & \{ \Gamma \vald \} \\
(\Gamma \Vdash \vec{F}, F_0 = \vec{G}, G_0 :: \Delta, x : K) & = & (\Gamma \Vdash \vec{F} = \vec{G} :: \Delta) \\
& & \cup (\Gamma \Vdash F_0 = G_0 : \{ \vec{F} / \Delta \} K)
\addtocounter{equation}{1}\end{eqnarray*}

\paragraph{Rules of Deduction}

The rules of deduction of $\TFk$ look exactly the same as those of TF, as given in Fig.~\ref{fig:TF}.  The rules (ctxt), (var) and (var\_eq) of course use the new definitions of the defined judgement forms $\Gamma \Vdash K \kind$, $\Gamma \Vdash \vec{F} :: \Delta$ and $\Gamma \Vdash \vec{F} = \vec{G} :: \Delta$.

Object theories are declared in $\TFk$ in the same way as in TF: we make a number of \emph{constant declarations} $c : (\Delta)T$, which has the effect of introducing the rules (const) and (const\_eq),
and \emph{equation declarations} $(\Delta)(M = N : T)$, which has the effect of introducing the rule (eq), as given in Section \ref{declarations}.  Again, in $\TFk$ these rules use the new definitions of the defined judgement forms.

\paragraph{Metatheory}

All the properties of TF we proved in Section \ref{section:metatheory} hold in $\TFk$ too.  The proofs follow the same pattern; we have indicated in Appendix \ref{appendix:A} the places where the details differ.

\subsection{Translations between $\mathrm{TF}$ and $\TFk$}

The systems TF and $\TFk$ are equivalent, in the following sense.  Given any derivable judgement in $\TFk$, erasing the kind labels on variables gives a derivable judgement in TF.  Conversely, given any derivable judgement in TF, there is a way of filling in the kind labels on the variables to yield a derivable judgement in TF; further, the choice of kind labels is unique up to equality in $\TFk$.

This fact is very convenient when working with lambda-free logical frameworks, as it allows us to switch between TF and $\TFk$ more or less at will, effectively treating them as if they were the same system.

In this section, we shall formally establish the equivalence of TF and $\TFk$ by defining translations between the two.

The translation from $\TFk$ to TF consists simply of erasing the kind labels:
\begin{df}
For every entity (object, abstraction, abstraction sequence, kind, context, or judgement) $X$ in $\TFk$, let $|X|$ denote the entity obtained by erasing the kind labels on the bound variables in abstractions.

Given a type theory specification $\mathcal{T}$ in $\TFk$, let $|\mathcal{T}|$ denote the type theory specification in TF formed by erasing the kind labels on the bound variables in abstractions within the declarations of $\mathcal{T}$.
\end{df}

It is straightforward to show that this translation is sound:
\begin{thm}
\label{thm:TFktoTF}
Let $\mathcal{T}$ be a type theory specification in $\TFk$, and let $J$ be a judgement that is derivable under $\mathcal{T}$.  Then $|J|$ is a derivable judgement in TF under the type theory specification $|\mathcal{T}|$.
\end{thm}

\begin{pf}
The proof consists of observing that the image of a primitive rule of deduction in $\TFk$ under $|\,|$ is a primitive rule of deduction in TF, and the image of any of the rules introduced by $\mathcal{T}$ under $|\,|$ is a rule introduced by $|\mathcal{T}|$.
\end{pf}

\pagebreak

Defining the translation in the other direction is harder.  We shall define the translation `$\lbl$' from TF to $\TFk$, which fills in the kind labels on the bound variables.  Whenever we encounter an object of the form $x[\cdots, [y_1, \ldots, y_n]M, \cdots]$, we discover the kinds of $y_1$, \ldots, $y_n$ by looking up the kind of $x$ in the current context.  Similarly, we handle objects of the form $c[\cdots]$ by looking up the kind of $c$ in the specification.

Let us say that an object, abstraction or abstraction sequence $X$ in TF is \emph{defined} relative to the specification $\mathcal{T}$ and context $\Gamma$ if and only if every constant that occurs in $X$ is declared in $\mathcal{T}$, and every free variable in $X$ is declared in $\Gamma$.  Let us also say that a context $\Delta \equiv x_1 : K_1, \ldots, x_n : K_n$ is \emph{defined} relative to $\Gamma$ and $\mathcal{T}$ if and only if, for each $i$, $K_i$ is defined relative to the context $\Gamma, x_1 : K_1, \ldots, x_{i-1} : K_{i-1}$ and $\mathcal{T}$.  Let us say that a judgement $\Gamma \vdash J$ is \emph{defined} relative to $\mathcal{T}$ if and only if $\Gamma$ is defined relative to $\mathcal{T}$, every constant that occurs in $J$ is declared in $\mathcal{T}$, and every free variable in $J$ is declared in $\Gamma$.

Let us say that the specification $\mathcal{T}$ is \emph{consistent} if and only if:
\begin{itemize}
\item
for each constant declaration $c : K$, the kind $K$ is defined relative to the empty context and $\mathcal{T}$;
\item
for each equation declaration $(\Delta)(M = N : T)$, the context $\Delta$ is defined relative to $\mathcal{T}$, and $M$, $N$ and $T$ are defined relative to $\Delta$ and $\mathcal{T}$.
\end{itemize}

Now, given a consistent specification $\mathcal{T}$ in TF, we shall define the following.
\begin{itemize}
\item
For every context $\Gamma$ defined relative to $\mathcal{T}$, and every object $M$ defined relative to $\mathcal{T}$ and $\Gamma$, an object $\lbl_\Gamma(M)$ in $\TFk$.
\item
For every abstraction $F$ and kind $K$ of the same arity defined relative to $\Gamma$ and $\mathcal{T}$, an abstraction $\lbl_\Gamma^K(F)$ in $TFk$.
We think of $K$ as the intended kind of $F$.
\item
For every abstraction sequence $\vec{F}$ and context $\Delta$ of the same arity defined relative to $\Gamma$ and $\mathcal{T}$, an abstraction sequence $\lbl_\Gamma^\Delta(\vec{F})$ in $\TFk$.
We think of $\Delta$ as giving the intended kinds of the abstractions $\vec{F}$.
\item
For every kind $K$ defined relative to $\Gamma$ and $\mathcal{T}$, a kind $\lbl_\Gamma(K)$ in $\TFk$.
\item
For every context $\Gamma$ defined relative to $\mathcal{T}$, a context $\lbl(\Gamma)$ in $\TFk$.
\item
For every judgement $J$ defined relative to $\mathcal{T}$, a judgement $\lbl(J)$ in $\TFk$.
\end{itemize}
The definition is as follows.
\begin{eqnarray*}
\lbl_\Gamma(c\vec{F}) & \equiv & c[\lbl_\Gamma^\Delta(\vec{F})] & ($c : (\Delta)T$ declared in $\mathcal{T}$) \\
\lbl_\Gamma(x\vec{F}) & \equiv & x[\lbl_\Gamma^\Delta(\vec{F})] & ($x : (\Delta)T$ declared in $\Gamma$) \\
& & \\
\lbl_\Gamma^T(M) & \equiv & \lbl_\Gamma(M) \\
\lbl_\Gamma^{(x:K)K'}([x]F) & \equiv & [x : \lbl_\Gamma(K)] \lbl_{\Gamma, x :K}^{K'}(F) \\
& & \\
\lbl_\Gamma^{\langle \rangle}(\langle \rangle) & \equiv & \langle \rangle \\
\lbl_\Gamma^{\Delta, x:K}(\vec{F}, G) & \equiv & \lbl_\Gamma^\Delta(\vec{F}), \lbl_\Gamma^{\{\vec{F} / \Delta \}K}(G) \\
& & \\
\lbl_\Gamma(\Type) & \equiv & \Type \\
\lbl_\Gamma(\El{M}) & \equiv & \El{\lbl_\Gamma(M)} \\
\lbl_\Gamma((x:K)K') & \equiv & (x : \lbl_\Gamma(K)) \lbl_{\Gamma, x : K}(K') \\
& & \\
\lbl(\langle \rangle) & \equiv & \langle \rangle \\
\lbl(\Gamma, x : K) & \equiv & \lbl(\Gamma), \lbl_\Gamma(K) \\
& & \\
\lbl(\Gamma \vald) & \equiv & \lbl(\Gamma) \vald \\
\lbl(\Gamma \vdash M : T) & \equiv & \lbl(\Gamma) \vdash \lbl_\Gamma(M) : \lbl_\Gamma(T) \\
\lbl(\Gamma \vdash M = N : T) & \equiv & \lbl(\Gamma) \vdash \lbl_\Gamma(M) = \lbl_\Gamma(N) : \lbl_\Gamma(T)
\addtocounter{equation}{1}\end{eqnarray*}

Given a consistent specification $\mathcal{S}$ in TF, let $\lbl(\mathcal{S})$ be the following type theory specification in $\TFk$.
\begin{itemize}
\item
For every constant declaration $c : K$ in $\mathcal{S}$, declare $c : \lbl_{\langle \rangle}(K)$.
\item
For every equation declaration $(\Delta)(M = N : T)$ in $\mathcal{S}$, declare \linebreak $(\lbl(\Delta))(\lbl_\Delta(M) = \lbl_\Delta(N) : \lbl_\Delta(T))$.
\end{itemize}

We can show that this translation is sound after proving a number of lemmas.

\begin{lm}
\label{lm:lblweak}
If $M$ is defined relative to both $\Gamma$ and $\Delta$, 
and $\Gamma$ and $\Delta$ agree on every free variable in $M$, then $\lbl_\Gamma(M) \equiv \lbl_\Delta(M)$.  In particular, if $M$ is defined relative to $\Gamma$ and $\Gamma \subseteq \Delta$, then $\lbl_\Gamma(M) \equiv \lbl_\Delta(M)$.
\end{lm}

\begin{pf}
An easy induction on $M$.
\end{pf}

\pagebreak

\begin{lm}
\label{lm:lblsub}
For each of the following equations, if the left-hand side is defined then so is the right-hand side, in which case the two are equal.
\begin{eqnarray*}
 \{ \lbl_\Gamma^K(F) / x \} \lbl_{\Gamma, x : K, \Delta}(X) & \equiv & \lbl_{\Gamma, \{F/x\}\Delta}(\{F/x\}X) \\
\{ \lbl_\Gamma^K(F) / x \} \lbl_{\Gamma, x : K, \Delta}^{K'}(G) & \equiv & \lbl_{\Gamma, \{F/x\}\Delta}^{\{F/x\}K'}(\{F/x\}G) \\
\{ \lbl_\Gamma^K(F) / x \} \lbl_{\Gamma, x : K, \Delta}^\Theta(\vec{G}) & \equiv & \lbl_{\Gamma, \{F/x\}\Delta}^{\{F/x\}\Theta}(\{F/x\}\vec{G}) \\
\addtocounter{equation}{1}\end{eqnarray*}
where $X$ is an object, kind or context.
\end{lm}

\begin{pf}
The five equations are proved simultaneously by a double induction on the arity of $K$, then the size of $X$, $G$ or $\vec{G}$.  We give the calculation for one case, the case where $X$ is an object of the form $x \vec{G}$.  Let $K \equiv (\Theta)T$ and $F \equiv [\dom \Theta]N$.
\begin{eqnarray*}
\{ \lbl(F) / x \} \lbl(x \vec{G}) & \equiv & \lbl(F) \bullet \{ \lbl(F) / x \} \lbl(\vec{G}) \\
& \equiv & \lbl(F) \bullet \lbl(\{F / x \} \vec{G}) & (i.h. on $X$) \\
& \equiv & \{ \lbl(\{F/x\} \vec{G}) / \Theta \} \lbl(N) \\
& \equiv & \lbl(\{\{F/x\} \vec{G} / \Theta\} N) & (i.h. on arity) \\
& \equiv & \lbl(\{F/x\}x\vec{G}) \qed
\addtocounter{equation}{1}\end{eqnarray*}
\end{pf}

The following lemma shows how we can change the subscript and superscript on an abstraction $\lbl_\Gamma^K(F)$.  Roughly, it can be read as: if $\lbl(\Gamma) = \lbl(\Gamma')$ and $\lbl(K) = \lbl(K')$, then $\lbl_\Gamma^K(F) = \lbl_{\Gamma'}^{K'}(F)$.

\begin{lm}
\label{lm:7}
The following rule of deduction is admissible in $\TFk$.
\[ \begin{prooftree}
\begin{array}{c}
\lbl_{\langle \rangle}(\Gamma) \Vdash \lbl_\Gamma^K(F) : \lbl_\Gamma(K) \\
\Vdash \lbl_{\langle \rangle}(\Gamma) = \lbl_{\langle \rangle}(\Gamma')
\qquad
\lbl_{\langle \rangle}(\Gamma) \Vdash \lbl_\Gamma(K) = \lbl_{\Gamma'}(K')
\end{array}
\justifies
\lbl_{\langle \rangle}(\Gamma) \Vdash \lbl_\Gamma^K(F) = \lbl_{\Gamma'}^{K'}(F) : \lbl_\Gamma(K)
   \end{prooftree} \]
\end{lm}

\begin{pf}
We prove that this rule and the following two are admissible.
\begin{equation}
 \begin{prooftree}
    \begin{array}{c}
     \lbl_{\langle \rangle}(\Gamma) \Vdash \lbl_\Gamma(M) : \lbl_\Gamma(T) \\
\Vdash \lbl_{\langle \rangle}(\Gamma) = \lbl_{\langle \rangle}(\Gamma')
    \end{array}
\justifies
\lbl_{\langle \rangle}(\Gamma) \Vdash \lbl_\Gamma(M) = \lbl_{\Gamma'}(M) : \lbl_\Gamma(T)
   \end{prooftree}
\label{eq:rule2}
\end{equation}
\[ \begin{prooftree}
    \begin{array}{c}
     \lbl_{\langle \rangle}(\Gamma) \Vdash \lbl_\Gamma^\Theta(\vec{F}) :: \lbl_\Gamma(\Theta) \\
\Vdash \lbl_{\langle \rangle}(\Gamma, \Theta) = \lbl_{\langle \rangle}(\Gamma', \Theta')
    \end{array}
\justifies
\lbl_{\langle \rangle}(\Gamma) \Vdash \lbl_{\Gamma}^{\Theta}(\vec{F}) = \lbl_{\Gamma'}^{\Theta'}(\vec{F}) :: \lbl_{\Gamma}(\Theta)
   \end{prooftree} \]

 The three rules are proved admissible simultaneously by induction on the size of $\lbl_\Gamma^K(F)$, $\lbl_\Gamma(M)$ and $\lbl_\Gamma^\Theta(\vec{F})$.  We give here the details for the case for (\ref{eq:rule2}) where $M$ has the form $x \vec{F}$.

Let $x$ have kind $(\Theta)S$ in $\Gamma$ and $(\Theta') S'$ in $\Gamma'$.  We are given that \linebreak $\lbl(\Gamma) \vdash \lbl_\Gamma(M) : \lbl(T)$.  Therefore, by Generation,
\[ \lbl(\Gamma) \Vdash \lbl_\Gamma^\Theta(\vec{F}) :: \lbl(\Theta) \qquad \lbl(\Gamma) \Vdash \{ \lbl_\Gamma^\Theta(\vec{F}) / \Theta \} \lbl(S) = \lbl(T) \enspace . \]
The induction hypothesis gives
\[ \lbl(\Gamma) \Vdash \lbl_\Gamma^\Theta(\vec{F}) = \lbl_{\Gamma'}^{\Theta'}(\vec{F}) :: \lbl(\Theta) \enspace . \]
and the desired result follows by (var\_eq) and (conv\_eq).
\end{pf}

\begin{lm}
\label{lm:8}
Let $T$ be an $n$-good declaration in $\TFk$.
The following rules of deduction are admissible in $\TFk$.
\[ \begin{array}{rc}
(\lbl\_\mathrm{seq}) & \begin{prooftree}
    \lbl(\Gamma \Vdash \vec{F} :: \Theta)
\justifies
\lbl_{\langle \rangle}(\Gamma) \Vdash \lbl_\Gamma^\Theta(\vec{F}) :: \lbl_\Gamma(\Theta)
   \end{prooftree} \\
& \\
(\lbl\_\mathrm{seqeq}) &
\begin{prooftree}
 \lbl(\Gamma \Vdash \vec{F} = \vec{G} :: \Theta)
\justifies
\lbl_{\langle \rangle}(\Gamma) \Vdash \lbl_\Gamma^\Theta(\vec{F}) = \lbl_\Gamma^\Theta(\vec{G}) :: \lbl_\Gamma(\Theta)
\end{prooftree} \end{array} \]
where $\Gamma$, $\vec{F}$, $\vec{G}$ and $\Theta$ are of order $\leq n$.
\end{lm}

\begin{pf}
We first prove the following two rules are admissible.
\[ \begin{array}{rc}
(\lbl\_\mathrm{abs}) & \begin{prooftree}
    \lbl(\Gamma \Vdash F : K)
\justifies
\lbl_{\langle \rangle}(\Gamma) \Vdash \lbl_\Gamma^K(F) : \lbl_\Gamma(K)
   \end{prooftree} \\
& \\
(\lbl\_\mathrm{abseq}) & \begin{prooftree}
 \lbl(\Gamma \Vdash F = G : K)
\justifies
\lbl_{\langle \rangle}(\Gamma) \Vdash \lbl_\Gamma^K(F) = \lbl_\Gamma^K(G) : \lbl_\Gamma(K)
\end{prooftree} \end{array} \]
For the first of these rules, if $K \equiv (\Theta)T$ and $F \equiv [\dom \Theta]M$, then the premise is $\lbl(\Gamma), \lbl(\Theta) \vdash \lbl(M) : \lbl(T)$, and the conclusion is
\[ (\lbl(\Gamma) \Vdash \lbl(\Theta) = \lbl(\Theta)) \cup \{ \lbl(\Gamma), \lbl(\Theta) \vdash \lbl(M) : \lbl(T) \} \]
which follows using Context Validity and (ref).  The proof for the second rule is similar.

The rules ($\lbl$\_seq) and ($\lbl$\_seqeq) are each proved admissible by induction on the length of $\vec{F}$.  We give the details for the rule ($\lbl$\_seqeq) where the length of $\vec{F}$ is greater than 0.
Suppose now that $\vec{F} \equiv \vec{F_0}, F_1$; $\vec{G} \equiv \vec{G_0}, G_1$; and $\Theta \equiv \Theta_0, x : K_1$.  The premises are
\[ \lbl(\Gamma \Vdash \vec{F_0} = \vec{G_0} :: \Theta_0) \cup  \lbl(\Gamma \Vdash F_1 = G_1 : \{ \vec{F_0} / \Theta_0 \} K_1) \]
and the conclusion is
\begin{eqnarray*}
& &  (\lbl(\Gamma) \Vdash \lbl(\vec{F_0}) = \lbl(\vec{G_0}) :: \lbl(\Theta)) \\
& \cup & (\lbl(\Gamma) \Vdash \lbl_\Gamma^{\{\vec{F_0} / \Theta_0\}K_1}(F_1) = \lbl_\Gamma^{\{\vec{G_0} / \Theta_0\}K_1}(G_1) : \lbl_\Gamma(\{\vec{F_0} / \Theta_0\}K_1) \enspace .
\addtocounter{equation}{1}\end{eqnarray*}
This follows, using the induction hypothesis, the rule ($\lbl$\_abseq) and Lemma \ref{lm:7}, once we have shown
\[ \lbl(\Gamma) \Vdash \lbl(\{\vec{F_0} / \Theta_0\} K_1) = \lbl(\{\vec{G_0} / \Theta_0\} K_1) \enspace . \]
By Lemma \ref{lm:lblsub}, this is
\[ \lbl(\Gamma) \Vdash \{ \lbl(\vec{F_0}) / \Theta_0 \} \lbl(K_1) = \{\lbl(\vec{G_0}) / \Theta_0 \} \lbl(K_1) \]
which is obtainable using Functionality.
\end{pf}

\begin{thm}
\label{thm:TFtoTFk}
Let $\mathcal{S}$ be an orderable $n$-good type theory specification in TF in which every declaration has order $\leq n$.  Assume we have declared $\mathcal{S}$ in TF and $\lbl(\mathcal{S})$ in $\TFk$.  Then, for every judgement $J$ derivable in TF with context of order $\leq n$, the judgement $\lbl(J)$ is derivable in $\TFk$.
\end{thm}

\begin{pf}
Let $\prec$ be the given order on $\mathcal{S}$.  For each declaration $\delta$ in $\mathcal{S}$, let $\mathcal{S}_\delta$ be the set of declarations $\delta'$ such that $\delta' \prec \delta$.  We prove the following simultaneously by $\prec$-induction on $\delta$:
\begin{enumerate}
\item
$\lbl(\mathcal{S}_\delta)$ is an orderable $n$-good specification in $\TFk$.
\item If $J$ is derivable in TF under $\mathcal{S}_\delta$, and $J$ has context of order $\leq n$, then $\lbl(J)$ is derivable under $\lbl(\mathcal{S}_\delta)$ in $\TFk$.
\end{enumerate}

The proof of 2 is by a straightforward induction on the derivation of $J$.  The cases (var), (const), (eq) all make use of the first rule in Lemma \ref{lm:8}; the cases (var\_eq) and (const\_eq) make use of the second rule in that lemma.
\end{pf}

Thus, our translations between TF and $\TFk$ are sound.  It is also easy to show that the mapping $|\;|$ is an exact left inverse to $\lbl$:
\begin{thm}
\[ |\lbl_\Gamma(X)| \equiv X \qquad
|\lbl_\Gamma^K(F)| \equiv F \qquad
|\lbl_\Gamma^\Delta(\vec{F})| \equiv \vec{F} \]
where $X$ is an object, kind or context.
\end{thm}

\begin{pf}
 An easy induction on $X$, $F$ and $\vec{F}$.
\end{pf}

The mapping $\lbl$ is \emph{not} a left inverse to $|\;|$ up to syntactic identity.  For example,
\[ \lbl_{A:\Type,B:\Type,C:\Type}^{(x : \El{A})\El{C}}(|[x : \El{B}]x|) \equiv [x : \El{A}]x \enspace . \]
However, on the well-typed objects, abstractions and kinds, $\lbl$ is a left inverse to $|\;|$ up to \emph{equality in $\TFk$}, in the following sense.

\pagebreak

\begin{thm}
Let $T$ be an $n$-good declaration in $\TFk$, and $\Gamma$, $F$, $K$, $\Delta$ have order $\leq n$.
\begin{enumerate}
 \item 
If $\Gamma \vdash M : T$ then $\Gamma \vdash M = \lbl_{|\Gamma|}(|M|) : T$.
\item
If $\Gamma \Vdash F : K$ then $\Gamma \Vdash F = \lbl_{|\Gamma|}^{|K|}(|F|) : K$.
\item
If $\Gamma \Vdash \vec{F} :: \Delta$ then $\Gamma \Vdash \vec{F} = \lbl_{|\Gamma|}^{|\Delta|}(|\vec{F}|) :: \Delta$.
\item
If $\Gamma \Vdash K \kind$ then $\Gamma \Vdash K = \lbl_{|\Gamma|}(|K|)$.
\item
If $\Gamma, \Delta \vald$ then $\Gamma \Vdash \Delta = \lbl_{|\Gamma|}(|\Delta|)$.
\end{enumerate}
\end{thm}

\begin{pf}
Let $l(X)$ denote the length of an expression $X$.
The five parts are proven simultaneously by induction on $l(\Gamma)+l(M)$, $l(\Gamma)+l(F)$, $l(\Gamma)+l(\vec{F})$, $l(\Gamma)+l(K)$, and $l(\Gamma)+l(\Delta)$.  We give here the details of the first two parts.
\begin{enumerate}
 \item 
Suppose $M \equiv x \vec{F}$, where $x : (\Theta)S \in \Gamma$.  By Generation,
\[ \Gamma \Vdash \vec{F} :: \Theta, \qquad \Gamma \Vdash \{ \vec{F} / \Theta \} S = T \enspace .\]
Therefore,
\begin{eqnarray*}
\quad \Gamma & \Vdash & \vec{F} = \lbl_{|\Gamma|}^{|\Theta|}(|\vec{F}|) :: \Theta & (i.h.) \\
\therefore \Gamma & \vdash & x \vec{F} = x \left[ \lbl_{|\Gamma|}^{|\Theta|} ( | \vec{F} | ) \right] : \{ \vec{F} / \Theta \} S & (var\_eq) \\
\therefore \Gamma & \vdash & x \vec{F} = x \left[ \lbl_{|\Gamma|}^{|\Theta|} ( | \vec{F} | ) \right] : T & (conv\_eq)
\addtocounter{equation}{1}\end{eqnarray*}
The case $M \equiv c \vec{F}$ is similar.
\item
Let $K \equiv (\Theta)T$ and $F \equiv [\Theta']M$.  We are given that
\[ \Gamma \Vdash \Theta = \Theta', \qquad \Gamma, \Theta \vdash M : T \enspace . \]
We must show that $\Gamma \Vdash [\Theta']M = [\lbl_{|\Gamma|}(|\Theta|)]\lbl_{|\Gamma|,|\Theta|}(|M|) : (\Theta) T$.  The induction hypothesis gives us that $\Gamma, \Theta \vdash M = \lbl_{|\Gamma|,|\Theta|}(|M|) : T$; it remains to show
\[ \Gamma \Vdash \Theta = \Theta' \enspace . \]
The induction hypothesis gives us that $\Gamma \Vdash \Theta' = \lbl_{|\Gamma|}(|\Theta'|)$; and $\Gamma \Vdash \Theta = \lbl_{|\Gamma|}(|\Theta|)$; it is thus sufficient to show
\[ \Gamma \Vdash \lbl_{|\Gamma|}(|\Theta|) = \lbl_{|\Gamma|}(|\Theta'|) \enspace . \]
Well,
\begin{eqnarray*}
\quad |\Gamma| & \Vdash_{\mathrm{TF}} & |\Theta| = |\Theta'| & (Theorem \ref{thm:TFktoTF}) \\
\therefore \lbl_{\langle \rangle}(|\Gamma|) & \Vdash_{\mathrm{TF}} & \lbl_{|\Gamma|}(|\Theta|) = \lbl_{|\Gamma|}(|\Theta'|) & (Theorem \ref{thm:TFtoTFk}) \\
& \Vdash_{\mathrm{TF}} & \Gamma = \lbl_{\langle \rangle}(|\Gamma|) & (i.h.)
\addtocounter{equation}{1}\end{eqnarray*}
and the result follows by Context Conversion.
\end{enumerate}
Parts 3--5 are proven similarly.
\end{pf}

We have thus established sound translations $|\,|$ and $\lbl$ between TF and $\TFk$ which are inverses of one another up to the appropriate notion of equality.

\section{Embedding $\mathrm{TF}$ in $\mathrm{LF}$}
\label{section:embed}

\begin{figure}
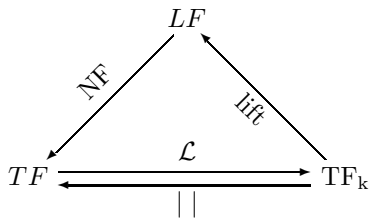

\[ \begin{diagram}
& & LF & &  \\
& \ldTo^{\NF} & & \luTo_{\lft} & \\
TF & & \pile{\rTo^{\mathcal{L}} \\ \lTo_{|\;|}} & & \TFk
\end{diagram} \]
\caption{Translations between Logical Frameworks}
\label{fig:trans}
\end{figure}

Lambda-free frameworks can often be embedded within existing traditional logical frameworks; that is, given a traditional logical framework $F$, we can often construct a lambda-free framework (its \emph{core}) that is, in some sense, isomorphic to a subsystem of $F$.  More precisely, we can construct a lambda-free framework $L$ and define translations 
\[ \NF : F \rightarrow L, \qquad \lift : L \rightarrow F \enspace . \]
These translations are sound, and $\NF$ is a left inverse to `$\lift$' up to identity ($\alpha$-conversion).  That is, we have the following properties:
\begin{enumerate}
\item
For every derivable judgement $J$ in $L$, $\lift(J)$ is derivable in $F$.
\item
For every derivable judgement $J$ in $F$, $\NF(J)$ is derivable in $L$.
\item
For every typable expression $X$ in $L$, $\NF(\lift(X)) \equiv X$.
\end{enumerate}
In many cases (particularly when $F$ allows $\eta$-conversion) we have in addition that $\NF$ is a right inverse to $\lift$ up to the equality judgements of $F$:
\begin{enumerate}
\setcounter{enumi}{3}
\item
For every typable expression $X$ in $F$, the equality $\lift(\NF(X)) = X$ is derivable in $F$.
\end{enumerate}
We can think of $F$ as picking out, from each equivalence class of the expressions of $F$ modulo $\beta \eta$-convertibility, a unique representative: the $\beta$-normal, $\eta$-long form.

Establishing the above properties of the translations is not easy; it usually involves proving fairly strong properties of $L$ and $F$.  However, once this one-time cost has been paid, we can then use the translations to prove various properties of $F$ more easily.  It is often the case that it is easier to establish a given metatheoretic property for $L$ than for $F$.  Once it has been proven to hold in $L$, the result can then be `lifted' to $F$; that is, we can derive the corresponding result for $F$ using the properties of the translations.

In this section, we shall show how TF can be embedded in this fashion within the framework LF
introduced in \cite{luo:car}, a Church-typed version of Martin-L\"{o}f's logical framework.  It will prove to be very advantageous that we have two different versions of TF; we shall define translations from $\TFk$ to LF, and from LF to TF, as shown in Figure \ref{fig:trans}.

\subsection{The Framework $\mathrm{LF}$}

The framework \LF~ \cite{luo:car} is a Church-typed version of Martin-L\"of's logical framework\footnote{The framework here called LF should not be confused with the Edinburgh Logical Framework \cite{hhp:lf}, which is also often referred to as LF.}.  \LF~ deals with \emph{objects} and \emph{kinds}, given by the following grammar:
\begin{eqnarray*}
 \mbox{Kind}~ K & ::= & \Type \mid \El{k} \mid (x : K) K \\
\mbox{Object}~ k & ::= & x \mid c \mid [x : K] k \mid kk
\addtocounter{equation}{1}\end{eqnarray*}
where $x$ is a variable and $c$ a constant.  There are five judgement forms in LF:
\begin{itemize}
 \item $\Gamma \vald$, which denotes that $\Gamma$ is a valid context;
\item $\Gamma \vdash K \kind$, which denotes that $K$ is a kind under $\Gamma$;
\item $\Gamma \vdash k : K$, which denotes that $k$ is an object of kind $K$ under $\Gamma$;
\item $\Gamma \vdash k = k' : K$, which denotes that $k$ and $k'$ are equal objects of kind $K$ under $\Gamma$;
\item $\Gamma \vdash K = K'$, which denotes that $K$ and $K'$ are equal kinds under $\Gamma$.
\end{itemize}
A type theory is specified in LF by giving a set of \emph{constant declarations} $c : K$, and a set of \emph{computation rules}
\[ k = k' : K \mbox{ for } k_1 : K_1, \ldots, k_n : K_n \enspace . \]

We shall make use of the following abbreviations when working with LF.  Let $\Delta$ be the context $x_1 : K_1, \ldots, x_n : K_n$, and $\Delta'$ the context $x_1 : K_1', \ldots, x_n : K_n'$.  We shall write $\Gamma \Vdash \Delta = \Delta'$ for the $n$ judgements
\begin{eqnarray*}
 \addtocounter{equation}{1}
\Gamma & \vdash & K_1 = K_1', \\
\Gamma, x_1 : K_1 & \vdash & K_2 = K_2', \\
& \vdots & \\
\Gamma, x_1 : K_1, \ldots, x_{n-1} : K_{n-1} & \vdash & K_n = K_n'
\end{eqnarray*}
and we shall write $\Gamma \Vdash (k_1, \ldots, k_n) :: \Delta$ for the $n$ judgements
\[ \Gamma \vdash k_1 : K_1, \quad \Gamma \vdash k_2 : [k_1 / x_1] K_2, \quad \ldots, \quad \Gamma \vdash k_n : [k_1 / x_1, \ldots, k_{n-1} / x_{n-1}] K_n \enspace . \]

For the rules of deduction of LF, and how LF may be used to specify various object theories, we refer to Luo \cite{luo:car}.

We note that, as with TF, the judgements of the object theory are represented by the LF-judgements of the form
\begin{eqnarray*}
 x_1 : \El{A_1}, \ldots, x_n : \El{A_n} & \vdash & k : \El{B} \\
x_1 : \El{A_1}, \ldots, x_n : \El{A_n} & \vdash & k = k' : \El{B}
\addtocounter{equation}{1}\end{eqnarray*}
and these are judgements with first-order contexts.

We shall make use of the fact that LF satisfies \emph{Subject Reduction}:
\begin{quote}
 If $\Gamma \vdash k : K$ and $k \twoheadrightarrow_{\beta \eta} k'$, then $\Gamma \vdash k = k' : K$.
\end{quote}

\subsection{Translation from $\TFk$ to $\mathrm{LF}$}

We shall now define our translations between LF and the two versions of TF.  The mapping from $\TFk$ to LF, which we shall call `$\lft$', is almost trivial.  We map objects and abstractions to objects, kinds to kinds, contexts to contexts and judgements to judgements as follows.
\begin{eqnarray*}
\lft(x[F_1, \ldots, F_n]) & \equiv & x \lft(F_1) \cdots \lft(F_n) \\
\lft([\Delta]M) & \equiv & [\lft(\Delta)] \lft(M) \\
& & \\
\lft(\Type) & \equiv & \Type \\
\lft(\El{M}) & \equiv & \El{\lft(M)} \\
\lft((x : K) K') & \equiv & (x : \lft(K)) \lft(K') \\
& & \\
\lft(x_1 : K_1, \ldots, x_n : K_n) & \equiv & x_1 : \lft(K_1), \ldots, x_n : \lft(K_n) \\
& & \\
\lft(\Gamma \vald) & \equiv & \lft(\Gamma) \vald \\
\lft(\Gamma \vdash M : T) & \equiv & \lft(\Gamma) \vdash \lft(M) : \lft(T) \\
\lft(\Gamma \vdash M = N : T) & \equiv & \lft(\Gamma) \vdash \lft(M) = \lft(N) : \lft(T)
\addtocounter{equation}{1}\end{eqnarray*}

It is relatively straightforward to establish that this translation is sound.

\begin{lm}
\label{lm:liftinst}
\[ [\lft(F) / x] \lft(N) \twoheadrightarrow_\beta \lft(\{F/x\}N) \]
\end{lm}

\begin{pf}
The proof is by a double induction on the arity of $F$ and $x$, then on the object $N$.
We give here the details for the case $N \equiv x \vec{G}$.  Let $F \equiv [\Delta]P$.
\begin{eqnarray*}
 [\lft(F) / x] x \lft(\vec{G}) & \equiv & \lft(F) [\lft(F) / x] \lft(\vec{G}) \\
& \equiv & ([\lft(\Delta)] \lft(P)) [\lft(F) / x] \lft(\vec{G}) \\
& \twoheadrightarrow & [[\lft(F) / x] \lft(\vec{G}) / \Delta] \lft(P) \\
& \twoheadrightarrow & [ \lft(\{F/x\}\vec{G}) / \Delta] \lft(P) & (i.h.) \\
& \twoheadrightarrow & \lft(\{ \{F/x\} \vec{G} / \Delta\} P) & (i.h.) \\
& \equiv & \lft(\{F/x\}N)
\addtocounter{equation}{1}\end{eqnarray*}
\end{pf}

\pagebreak

\begin{thm}
\label{thm:liftsound}
Suppose we have declared a type theory $T$ in $\TFk$, and the corresponding theory $\lft(T)$ in LF.
If $\mathcal{J}$ is a derivable judgement in $\TFk$, then $\lift(\mathcal{J})$ is derivable in LF.
\end{thm}

\begin{pf}
We first prove that the following rules of deduction are admissible in LF:
\[ (\lft\_\mathrm{abs}) \; \begin{prooftree}
\lft(\Gamma \Vdash F : K)
\justifies
\lft(\Gamma) \vdash \lft(F) : \lft(K)
\end{prooftree}
\] \[
(\lft\_\mathrm{abseq}) \; \begin{prooftree}
\lft(\Gamma \Vdash F = G : K)
\justifies
\lft(\Gamma) \vdash \lft(F) = \lft(G) : \lft(K)
\end{prooftree} \]
\[ (\lft\_\mathrm{seq}) \; \begin{prooftree}
\lft(\Gamma \Vdash \vec{F} :: \Delta) \quad \lft(\Gamma, \Delta \vald)
\justifies
\lft(\Gamma) \Vdash \lft(\vec{F}) :: \lft(\Delta)
\end{prooftree}
\] \[
(\lft\_\mathrm{seqeq}) \; \begin{prooftree}
\lft(\Gamma \Vdash \vec{F} = \vec{G} :: \Delta) \quad \lft(\Gamma, \Delta \vald)
\justifies
\lft(\Gamma) \Vdash \lft(\vec{F}) = \lft(\vec{G}) :: \lft(\Delta)
\end{prooftree} \]

%
The proof for ($\lft$\_seq) is by induction on the length of $\vec{F}$.

If the length is 0, both hypothesis and conclusion are that $\lft(\Gamma)$ is valid.

Suppose $\vec{F}$ is of length $n+1$, and the result holds for abstraction sequences of length $n$.  Let $\vec{F} \equiv \vec{F_0}, F_1$; and $\Delta \equiv \Delta_0, x:K_1$.  We are given that $\lft(\Gamma \Vdash \vec{F_0} :: \Delta_0)$ is derivable, hence so is $\lft(\Gamma) \Vdash \lft(\vec{F_0}) :: \lft(\Delta_0)$ by the induction hypothesis.  We also have
\[ \lft(\Gamma) \vdash \lft(F_1) : \lft(\{\vec{F_0} / \Delta_0\} K_1) \]
by part 1 and
\[ \lft(\Gamma), \lft(\Delta_0) \vdash \lft(K_1) \kind \]
by Kind Validity in LF.  This yields
\begin{eqnarray*}
\quad \lft(\Gamma) & \vdash & [\lft(\vec{F_0}) / \Delta_0] \lft(K_1) \kind & (substitution) \\
\therefore \lft(\Gamma) & \vdash & [\lft(\vec{F_0}) / \Delta_0] \lft(K_1) = \lft(\{\vec{F_0} / \Delta_0\} K_1) \\
& & &  (Subject Reduction, Lemma \ref{lm:liftinst}) \\
\therefore \lft(\Gamma) & \vdash & \lft(F_1) : [\lft(\vec{F_0}) / \Delta_0] \lft(K_1) & (conv)
\addtocounter{equation}{1}\end{eqnarray*}
as required.

The proof for ($\lft$\_seqeq) is similar, and the proofs for (lift\_abs) and (lift\_seq) are simple.
The theorem now follows by induction on the derivation of $\mathcal{J}$.
\end{pf}

\pagebreak

\subsection{Translation from $\mathrm{LF}$ to $\mathrm{TF}$}

The translation from LF to TF is more difficult to construct.  It consists of reducing every entity of LF to its $\beta$-normal, $\eta$-long form.

We must first assign arities to the entities of LF, to guide us during $\eta$-expansion.  We assign an arity to every kind of LF as follows:
\begin{eqnarray*}
\Ar(\Type) & \equiv & \mathbf{0} \\
\Ar(\El{k}) & \equiv & \mathbf{0} \\
\Ar((x:K_1)K_2) & \equiv & (\Ar(K_1)) \abs \Ar(K_2)
\addtocounter{equation}{1}\end{eqnarray*}
We now define an arity $\Ar_\Gamma(k)$ to some LF-contexts $\Gamma$ and LF-objects $k$ as follows:
\begin{itemize}
 \item If $x:K$ is an entry in $\Gamma$, then $\Ar_\Gamma(x) \equiv \Ar(K)$.
\item If $c$ has been declared with arity $K$, then $\Ar_\Gamma(c) \equiv \Ar(K)$.
\item If $\Ar_{\Gamma, x:K}(k)$ is defined, then $\Ar_\Gamma([x:K]k) \equiv (\Ar(K)) \abs \Ar_{\Gamma,x:K}(k)$.
\item If $\Ar_\Gamma(k)$ and $\Ar_\Gamma(k')$ is defined, and $\Ar_\Gamma(k)$ has the form
\[ \Ar_\Gamma(k) \equiv (\Ar_\Gamma(k')) \abs \beta \]
then $\Ar_\Gamma(kk') \equiv \beta$.
\end{itemize}
We shall say that an object $k$ is \emph{well-aritied} if $\Ar_\Gamma(k)$ is defined.  We shall only be able to map well-aritied objects into TF.  We can prove immediately that every object typable in LF is well-aritied.
\begin{prop}
In LF,
\begin{enumerate}
 \item if $\Gamma \vdash k : K$ then $\Ar_\Gamma(k) \equiv \Ar(K)$;
\item if $\Gamma \vdash k = k' : K$ then $\Ar_\Gamma(k) \equiv \Ar_\Gamma(k') \equiv \Ar(K)$;
\item if $\Gamma \vdash K = K'$ then $\Ar(K) \equiv \Ar(K')$.
\end{enumerate}
\end{prop}

\begin{pf}
The three statements are proven simultaneously by induction on the derivation of the premise.  We need to make use of the following two auxiliary facts, which are easy to prove:
\begin{enumerate}
 \item $\Ar([k/x]K) \equiv \Ar(K)$
\item If $\Ar_\Gamma(k) \equiv \Ar(K)$ and $\Ar_{\Gamma, x:K}(k')$ is defined, then we have \linebreak
$\Ar_\Gamma([k/x]k') \equiv \Ar_{\Gamma, x:K}(k')$. \qed
\end{enumerate}
\end{pf}

Given an object $k$ such that $\Ar_\Gamma(k) \equiv \alpha$, we define the $\alpha$-ary abstraction $\NF_\Gamma(k)$ in TF as follows.
\begin{eqnarray*}
\NF_\Gamma(x) & \equiv & x^\eta \\
\NF_\Gamma(c) & \equiv & c^\eta \\
\NF_\Gamma([x : K]k) & \equiv & [x] \NF_{\Gamma, x : K}(k) \\
\NF_\Gamma(kk') & \equiv & \NF_\Gamma(k) \bullet \NF_\Gamma(k')
\addtocounter{equation}{1}\end{eqnarray*}
where, in the first two clauses, $x$ has arity $\Ar_\Gamma(x)$ and $c$ has arity $\Ar_\Gamma(c)$.  In the third clause, $x$ has arity $\Ar_\Gamma(K)$.

\pagebreak

We extend the mapping NF to kinds, contexts and judgements as follows.
\begin{eqnarray*}
 \NF_\Gamma(\Type) & \equiv & \Type \\
\NF_\Gamma(\El{k}) & \equiv & \El{\NF_\Gamma(k)} \\
\NF_\Gamma((x:K)K') & \equiv & (x : \NF_\Gamma(K)) \NF_{\Gamma, x:K}(K') \\
& & \\
\NF_\Gamma(\langle \rangle) & \equiv & \langle \rangle \\
\NF_\Gamma(\Delta, x:K) & \equiv & \NF_\Gamma(\Delta), x : \NF_{\Gamma, \Delta}(K) \\
& & \\
\NF(\Gamma \vald) & = & \{ \NF_{\langle \rangle}(\Gamma) \vald \} \\
\NF(\Gamma \vdash K \kind) & = & (\NF_{\langle \rangle}(\Gamma) \Vdash \NF_\Gamma(K) \kind ) \\
\NF(\Gamma \vdash K = K') & = & (\NF_{\langle \rangle}(\Gamma) \Vdash \NF_\Gamma(K) = \NF_\Gamma(K')) \\
\NF(\Gamma \vdash k : K) & = & ( \NF_{\langle \rangle}(\Gamma) \Vdash \NF_\Gamma(k) : \NF_\Gamma(K) ) \\
\NF(\Gamma \vdash k = k' : K) & = & ( \NF_{\langle \rangle}(\Gamma) \Vdash \NF_\Gamma(k) = \NF_\Gamma(k') : \NF_\Gamma(K) )
\addtocounter{equation}{1}\end{eqnarray*}
Given a type theory specification $\mathcal{T}$ in LF, we define the type theory specification $\NF(\mathcal{T})$ in TF as follows.
\begin{itemize}
 \item For each declaration $c : K$ in $\mathcal{T}$, the declaration $c : \NF_{\langle \rangle}(K)$ is in $\NF(\mathcal{T})$.
\item For each declaration $(\Delta)(k = k' : K)$ in $\mathcal{T}$, the declaration \linebreak $(\NF_{\langle \rangle}(\Delta))(\NF_\Delta(k) = \NF_\Delta(k') : \NF_\Delta(K))$ is in $\NF(\mathcal{T})$.
\end{itemize}

The following results ensure that this translation is well-behaved and sound.
\begin{thm}$ $
\label{thm:NFsound}
 \begin{enumerate}
  \item Let $\Ar(K) \equiv \alpha$.  If $\NF_\Gamma(K)$ is defined, then it is an $\alpha$-ary kind.
\item Let $\Gamma \subseteq \Delta$.  If $\NF_\Gamma(X)$ is defined, then $\NF_\Delta(X)$ is defined, and
\[ \NF_\Delta(X) \equiv \NF_\Gamma(X) \enspace . \]
\item
Suppose $\Ar_\Gamma(k) \equiv \Ar(K)$.  Let $X$ be an LF-object, kind or context.  If $\NF_\Gamma(k)$ and $\NF_{\Gamma, x:K, \Delta}(X)$ are defined, then $\NF_{\Gamma, [k/x]\Delta}([k/x]X)$ is defined, and
\[ \NF_{\Gamma, [k/x]\Delta}([k/x]X) \equiv \{\NF_\Gamma(k) / x\} \NF_{\Gamma,x:K,\Delta}(X) \enspace . \]
\item
Let $T$ be a type-theory specification in LF, and suppose $\NF(T)$ is an $n$-good specification in TF.
If the judgement $\mathcal{J}$ is derivable in LF and has context of order $\leq n$, then $\NF(\mathcal{J})$ is defined and derivable in TF.
 \end{enumerate}
\end{thm}

\pagebreak

\begin{pf}
The first three parts are easily proven by an induction on $K$ and $X$ respectively.


The fourth part is proven by induction on the derivation of $\mathcal{J}$.  Most cases are straightforward, making use of the results proven in Section \ref{section:metatheory}.  We give here the details for the rule (beta).
\[ (\mathrm{beta}) \; \begin{prooftree}
\Gamma, x : K \vdash k' : K' \qquad \Gamma \vdash k : K
\justifies
\Gamma \vdash ([x:K]k')k = [k/x]k' : [k/x]K'
\end{prooftree} \]
By the induction hypothesis,
\begin{eqnarray*}
 \addtocounter{equation}{1}
\NF_{\langle \rangle}(\Gamma), x : \NF_\Gamma(K) & \Vdash & \NF_{\Gamma, x : K}(k') : \NF_{\Gamma, x:K}(K'); \\
\NF_{\langle \rangle}(\Gamma) & \Vdash & \NF_\Gamma(k) : \NF_\Gamma(K) \enspace .
\end{eqnarray*}
Now,
\begin{eqnarray*}
\NF_\Gamma(([x:K]k')k) & \equiv & ([x] \NF_{\Gamma, x:K}(k')) \bullet \NF_\Gamma(k) \\
& \equiv & \{ \NF_\Gamma(k) / x \} \NF_{\Gamma, x:K}(k') \\
& \equiv & \NF_\Gamma([k/x]k') & (part 3)
\addtocounter{equation}{1}\end{eqnarray*}
The Cut rule and (ref) give us
\begin{eqnarray*} 
\addtocounter{equation}{1}
\NF_{\langle \rangle}(\Gamma) & \Vdash & \{ \NF_\Gamma(k) / x \} \NF_{\Gamma, x:K}(k') = \{ \NF_\Gamma(k) / x \} \NF_{\Gamma, x:K}(k') \\
& : & \{ \NF_\Gamma(k) / x \} \NF_{\Gamma, x:K}(K')
\end{eqnarray*}
and, by part 3, this is the same judgement as
\[ \NF_{\langle \rangle}(\Gamma) \Vdash \NF_\Gamma(([x:K]k')k) = \NF_\Gamma([k/x]k') : \NF_\Gamma([k/x]K') \enspace . \]

\end{pf}

The translations we have established between our three systems are shown in Figure \ref{fig:trans}.  The triangles in this diagram commute in the sense given by the following theorem.

\begin{thm}
\label{thm:inverses}
Let $T$ be a type theory specification in LF, and suppose $\NF(T)$ is an orderable $n$-good type theory specification in TF.
\begin{enumerate}
\item
If $\Gamma \vdash k : K$ in LF, then
\[ \Gamma \vdash k = \lift\left(\mathcal{L}_{\NF_{\langle \rangle}(\Gamma)}^{\NF_\Gamma(K)}(\NF_\Gamma(k))\right) : K \enspace . \]
Similar results hold for kinds and contexts.
\item
If $\Gamma \vdash M : T$ in TF, then
\[ M \equiv \NF_{\lift(\mathcal{L}_{\langle \rangle}(\Gamma))}(\lift(\mathcal{L}_\Gamma(M))) \enspace . \]
Similar results hold for kinds and contexts.
\item
If $\Gamma \vdash M : T$ in $\TFk$, then
\[ \Gamma \vdash M = \mathcal{L}_{\NF_{\langle \rangle}(\lift(\Gamma))}(\NF_{\lift(\Gamma)}(\lift(M))) : T \enspace . \]
Similar results hold for kinds and contexts.
\end{enumerate}
\end{thm}

\begin{pf}$ $
\begin{enumerate}
\item
We prove the statement:
\begin{quote}
If $\Gamma \vdash k : K$ and $\Vdash \Gamma = \Delta$ in LF, then
\[ \Gamma \vdash k = \lift(\mathcal{L}_{\NF_{\langle \rangle}(\Gamma)}^{\NF_\Gamma(K)}(\NF_\Delta(k))) : K \enspace . \]
\end{quote}
We prove the statements simultaneously with similar statements for kinds and contexts by induction on size.
We give here the details for the case where $k$ is an abstraction.

Let $k \equiv [x : K_0]k'$, and $K \equiv (x : K_1) K_2$.  By Generation, we have
\[ \Gamma \vdash K_0 = K_1, \qquad \Gamma, x : K_1 \vdash k' : K_2 \enspace . \]
Now,
\begin{eqnarray*}
\lefteqn{\lift\left(\mathcal{L}_{\NF_{\langle \rangle}(\Gamma)}^{\NF_\Gamma(K)}(\NF_\Delta(k))\right)} \\
 & \equiv & \lift\left(\mathcal{L}_{\NF_{\langle \rangle}(\Gamma)}^{(x : \NF_\Gamma(K_1))\NF_{\Gamma, x:K_1}(K_2)}([x] \NF_{\Delta, x : K_0}(k'))\right) \\
& \equiv & \lift\left([x : \mathcal{L}_{\NF_{\langle \rangle}(\Gamma)}(\NF_\Gamma(K_1))] \mathcal{L}_{\NF_{\langle \rangle}(\Gamma, x : K_1)}^{\NF_{\Gamma, x :K_1}(K_2)}(\NF_{\Delta, x : K_0}(k'))\right) \\
& \equiv & [x : \lift(\mathcal{L}_{\NF_{\langle \rangle}(\Gamma)}(\NF_\Gamma(K_1)))] \lift\left(\mathcal{L}_{\NF_{\langle \rangle}(\Gamma, x : K_1)}^{\NF_{\Gamma, x : K_1}(K_2)}(\NF_{\Delta, x : K_0}(k'))\right)
\addtocounter{equation}{1}\end{eqnarray*}
Now, the induction hypothesis gives the two judgements
\begin{eqnarray*}
\Gamma & \vdash & \lift(\mathcal{L}_{\NF_{\langle \rangle}(\Gamma)}(\NF_\Gamma(K_1))) = K_1 \\
\Gamma, x : K_1 & \vdash & \lift\left(\mathcal{L}_{\NF_{\langle \rangle}(\Gamma, x : K_1)}^{\NF_{\Gamma, x : K_1}(K_2)}(\NF_{\Delta, x : K_0}(k'))\right) = k' : K_2 
\addtocounter{equation}{1}\end{eqnarray*}
from which the result follows.
\item
The proof is by induction on the object $M$.  Let $M \equiv z[\vec{F}]$, and let $z$ have kind $(\Delta)T$ relative to $\Gamma$.  Then
\begin{eqnarray*}
\NF_{\lift(\mathcal{L}_{\langle \rangle}(\Gamma))}(\lift(\mathcal{L}_\Gamma(M))) & \equiv & \NF_{\lift(\mathcal{L}_{\langle \rangle}(\Gamma))}(\lift(z[\mathcal{L}_{\langle \rangle}^\Delta(\vec{F})])) \\
& \equiv & \NF_{\lift(\mathcal{L}_{\langle \rangle}(\Gamma))}(z[\lift(\mathcal{L}_{\langle \rangle}^\Delta(\vec{F}))]) \\
& \equiv & z^\eta \bullet \NF_{\lift(\mathcal{L}_{\langle \rangle}(\Gamma))}(\lift(\mathcal{L}_{\langle \rangle}^\Delta(\vec{F}))) \\
& \equiv & z[\NF_{\lift(\mathcal{L}_{\langle \rangle}(\Gamma))}(\lift(\mathcal{L}_{\langle \rangle}^\Delta(\vec{F})))] \\
& \equiv & z[\vec{F}] & (i.h.)
\addtocounter{equation}{1}\end{eqnarray*}
\item
The proof is by induction on the object $M$.  Let $M \equiv x[\vec{F}]$, and let 
\[ \Gamma \equiv \Gamma_1, x:(\Delta)S, \Gamma_2 \enspace . \]
Then
\begin{eqnarray*}
\lefteqn{\mathcal{L}_{\NF_{\langle \rangle}(\lift(\Gamma))}(\NF_{\lift(\Gamma)}(\lift(M)))} \\
 & \equiv & \mathcal{L}_{\NF_{\langle \rangle}(\lift(\Gamma))}(\NF_{\lift(\Gamma)}(x \lift(\vec{F}))) \\
& \equiv & \mathcal{L}_{\NF_{\langle \rangle}(\lift(\Gamma))}(x^\eta \bullet \NF_{\lift(\Gamma)}(\lift(\vec{F}))) \\
& \equiv & \mathcal{L}_{\NF_{\langle \rangle}(\lift(\Gamma))}(x[\NF_{\lift(\Gamma)}(\lift(\vec{F}))]) \\
& \equiv & x\left[\mathcal{L}_{\NF_{\langle \rangle}(\lift(\Gamma))}^{\NF_{\lift(\Gamma_1)}(\lift(\Delta))}(\NF_{\lift(\Gamma)}(\lift(\vec{F})))\right] \\
& \equiv & x\left[\mathcal{L}_{\NF_{\langle \rangle}(\lift(\Gamma))}^{\NF_{\lift(\Gamma)}(\lift(\Delta))}(\NF_{\lift(\Gamma)}(\lift(\vec{F})))\right]
\addtocounter{equation}{1}\end{eqnarray*}
Now, by Generation, $\Gamma \Vdash \vec{F} :: \Delta$ and $\Gamma \Vdash \{\vec{F} / \Delta \} S = T$.  Hence, the induction hypothesis gives
\[ \Gamma \Vdash \vec{F} = \mathcal{L}_{\NF_{\langle \rangle}(\lift(\Gamma))}^{\NF_{\lift(\Gamma)}(\lift(\Delta))}(\NF_{\lift(\Gamma)}(\vec{F})) :: \Delta \]
from which the result follows. \qed
\end{enumerate}
\end{pf}

\subsection{Lifting Results}

Suppose we wish to establish a property of a framework, or of an object theory in a traditional framework $F$.  It is often the case that the property is more easily proven for a lambda-free framework $L$.  The result can then be `lifted' to $F$; that is, we can derive the result for $F$ easily from $L$, together with the properties of the translations between $L$ and $F$.

In Luo and Adams \cite{la:ssitfer}, we were working with a type theory declared in LF: an extension of the type theory UTT \cite{luo:car} with some new reduction rules.  It was found to be necessary to prove that type constructors are injective; that is, whenever $T : (K) \Type$ and $TA = TB$, then $A = B$.  We were not able to find a way to prove this result in LF directly; the obvious method requires using the Church-Rosser property for the new reduction relation, which is not known to hold.  However, the corresponding result in TF is almost trivial, and so we made use of this fact and lifted the result from TF to LF.  As an illustration of the process of lifting results, we repeat the details here.

\pagebreak

We seek to prove:
\begin{thm}[Injectivity of Type Constructors]
\label{thm:injtype}
Let $\mathcal{S}$ be a type theory specification in LF that has the property: for every equation declaration \linebreak $(\Delta)(M = N : T)$ in $\mathcal{S}$, $T$ has the form $\El{A}$ (that is, there are no equation declarations of the form $(\Delta)(M = N : \Type)$).  Further, suppose $\NF(\mathcal{S})$ is an orderable $n$-good specification in TF.  Let $c : (\Theta) \Type$ be a constant declaration in $\mathcal{S}$.  Then the following rule of deduction is admissible:
\[ \begin{prooftree}
\Gamma \vdash c \vec{A} = c \vec{B} : \Type
\justifies
\Gamma \Vdash \vec{A} = \vec{B} :: \Theta
\end{prooftree} \]
where $\Gamma$ has order $\leq n$.
\end{thm}

The corresponding result for TF is fairly easy to prove:
\begin{thm}
\label{thm:TFinj}
Let $\mathcal{S}$ be a type theory specification in TF that has the property: for every equation declaration $(\Delta)(M = N : T)$ in $\mathcal{S}$, $T$ has the form $\El{A}$.  Let $c : (\Theta) \Type$ be a constant declaration in $\mathcal{S}$.  Then the following rule of deduction is admissible:
\[ \begin{prooftree}
\Gamma \vdash c \vec{F} = c \vec{G} : \Type
\justifies
\Gamma \Vdash \vec{F} = \vec{G} :: \Theta
\end{prooftree} \]
\end{thm}

\begin{pf}
We shall prove the following statement.
\begin{quote}
If $\Gamma \vdash c \vec{F} = X : \Type$ or $\Gamma \vdash X = c \vec{F} : \Type$ is derivable, then $X$ has the form $c \vec{G}$, and $\Gamma \Vdash \vec{F} = \vec{G} :: \Theta$.
\end{quote}
The proof is by induction on the derivation of the premise.  Note that the last step in this derivation cannot be the use of an equation from $\mathcal{S}$.  All cases are straightforward.
\end{pf}

The result can now be `lifted' to LF.  We omit the sub- and superscripts on NF and $\lbl$ in the following proof.
\paragraph{Proof of Theorem \ref{thm:injtype}}
Let $\mathcal{S}$ satisfy the hypotheses of the theorem.  Suppose $\Gamma \vdash c \vec{A} = c \vec{B} : \Type$ is derivable in LF under $\mathcal{S}$.  By Theorem \ref{thm:NFsound},
\[ \NF(\Gamma) \vdash c \NF(\vec{A}) = c \NF(\vec{B}) : \Type \]
is derivable in TF under $\NF(\mathcal{S})$.  We note also that $\NF(\mathcal{S})$ satisfies the hypotheses of Theorem \ref{thm:TFinj}.  Therefore,
\begin{eqnarray*}
\NF(\Gamma) & \Vdash_{TF} & \NF(\vec{A}) = \NF(\vec{B}) :: \NF(\Theta) & (Theorem \ref{thm:TFinj}) \\
\therefore \lbl(\NF(\Gamma)) & \Vdash_{TF} & \lbl(\NF(\vec{A})) = \lbl(\NF(\vec{B})) :: \lbl(\NF(\Theta)) & (Theorem \ref{thm:TFtoTFk}) \\
\therefore \lift(\lbl(\NF(\Gamma))) & \Vdash_{LF} & \lift(\lbl(\NF(\vec{A}))) = \lift(\lbl(\NF(\vec{B}))) :: \lift(\lbl(\NF(\Theta))) \\
& & & (Theorem \ref{thm:liftsound})
\addtocounter{equation}{1}\end{eqnarray*}
We also have, by Theorem \ref{thm:inverses},
\begin{eqnarray*}
& \Vdash_{LF} & \Gamma = \lift(\lbl(\NF(\Gamma))) \\
\Gamma & \Vdash_{LF} & \vec{A} = \lift(\lbl(\NF(\vec{A}))) :: \Theta \\
\Gamma & \Vdash_{LF} & \vec{B} = \lift(\lbl(\NF(\vec{B}))) :: \Theta \\
\Gamma & \Vdash_{LF} & \Theta = \lift(\lbl(\NF(\Theta)))
\addtocounter{equation}{1}\end{eqnarray*}
It follows that
\[ \Gamma \Vdash_{LF} \vec{A} = \vec{B} :: \Theta \]
as required.

$ $

In contrast, the author has been unable to find a direct proof of this result in LF.

Here is a second example of how a result may be lifted from TF to LF.  Let $\mathcal{T}$ be a type theory specification in LF.  Roughly, we shall show that, if $\NF(\mathcal{T})$ is strongly normalising in TF, then $\mathcal{T}$ is strongly normalising in LF.

More strictly, assume we have declared $\mathcal{T}$ in LF and $\NF(\mathcal{T})$ in TF.  Suppose $\NF(\mathcal{T})$ is orderable and 1-good.  Let $\rightarrow_R$ be a reduction relation on the objects of LF, and let $\rightarrow_{R \beta \eta}$ be the union of $\rightarrow_R$ and framework-level $\beta$- and $\eta$-reduction:
\begin{eqnarray*}
 ([x:K]k)k' & \rightarrow_\beta & [k'/x]k \\
{[x:K]}kx & \rightarrow_\eta & k \enspace .
\addtocounter{equation}{1}\end{eqnarray*}
Define the relation $\rhd$ on the objects of TF as follows:
\begin{quote}
 $M \rhd N$ if and only if there exist LF-objects $a$, $b$ such that $\NF(a) = M$, $\NF(b) = N$, and $M \rightarrow_R N$.
\end{quote}
Then we have
\begin{thm}
Suppose that every object typable in TF is strongly \linebreak $\rhd$-normalising.  Then every object typable in LF is strongly $\rightarrow_{R \beta \eta}$-normalising.
\end{thm}

\begin{pf}
 Suppose $\Gamma \vdash a : A$ and 
\begin{equation}
 \label{eq:redseq}
a \rightarrow_{R \beta \eta} a_1 \rightarrow_{R \beta \eta} a_2 \rightarrow_{R \beta \eta} \cdots
\end{equation}
 is an infinite $\rightarrow_R$-reduction sequence starting with $a$.  Then $\NF(\Gamma) \vdash_{TF} \NF(a) : \NF(A)$, so $\NF(a)$ is strongly $\rhd$-normalisable.

Now, if $a_n \rightarrow_R a_{n+1}$, then $\NF(a_n) \rhd \NF(a_{n+1})$; and if $a_n \rightarrow_{\beta \eta} a_{n+1}$, then $\NF(a_n) \equiv \NF(a_{n+1})$.  So we have
\[ \NF(a) \trianglerighteq \NF(a_1) \trianglerighteq \NF(a_2) \trianglerighteq \cdots \enspace . \]
This sequence cannot contain an infinite number of $\rhd$-reductions; therefore, there must be some $n$ such that
\[ \NF(a_n) \equiv \NF(a_{n+1}) \equiv \NF(a_{n+2}) \equiv \cdots \]
and hence
\[ a_n \rightarrow_{\beta \eta} a_{n+1} \rightarrow_{\beta \eta} a_{n+2} \rightarrow_{\beta \eta} \cdots \enspace . \]
This contradicts the fact that LF is strongly $\rightarrow_{\beta \eta}$-normalising.
\end{pf}

It is often easier to prove that $\NF(\mathcal{T})$ is strongly $\rhd$-normalising than that $\mathcal{T}$ is strongly $\rightarrow_{R \beta \eta}$-normalising, because we do not have to consider how $\rightarrow_R$ and framework-level $\beta$- and $\eta$-reduction interact.

We have made use in this proof of the fact that LF is strongly $\beta \eta$-normalising under an arbitrary type theory specification.  This is not difficult to prove, but, to the best of the author's knowledge, a proof has not yet been published, and so we present one in Appendix \ref{appendix:lfsn}.

\section{Related Work}
\label{section:related}

Several lambda-free logical frameworks have appeared, independently, since the publication of Adams \cite{adams:thesis}.

\subsection{The Canonical Logical Framework}

The \emph{Canonical Logical Framework} (Canonical LF) \cite{hl:mmlf, lp:brtslf} is a subsystem of the Edinburgh Logical Framework (ELF) that deals only with objects in $\beta$-normal, $\eta$-long forms.  This framework uses an operation of \emph{hereditary substitution} $[M/x]_\alpha^mN$ which behaves similarly to TF's instantiation.  Their operation must be given a simple type $\alpha$, which plays a similar role to the arity in TF.

The Canonical LF is essentially the same system as the following subsystem of $\TFk$.  Let us say that a product kind $(x_1 : K_1, \ldots, x_n : K_n) T$ is \emph{small} iff the symbol $\Type$ does not occur in it, and \emph{large} otherwise.  We impose the following restrictions on TF:
\begin{itemize}
 \item every variable that appears in a judgement or constant declaration must have a small kind;
\item no equation declarations may be made.
\end{itemize}
This subsystem was the system named $\mathbf{SPar}(\omega)^-$ in \cite{adams:thesis}.  We can prove that $\TFk$ is conservative over this subsystem in a very strong sense:
\begin{thm}
Let $\mathcal{T}$ be a type theory specification containing no equation declarations, such that every variable in a constant declaration has a small kind.  Let $\mathcal{J}$ be a judgement in which every variable has a small kind.  If $\mathcal{J}$ is derivable in $\TFk$ under $\mathcal{T}$, then $\mathcal{J}$ is derivable under $\mathcal{T}$ in $\mathbf{SPar}(\omega)^-$.  In fact, every derivation of $\mathcal{J}$ in $\TFk$ is a derivation of $\mathcal{J}$ in $\mathbf{SPar}(\omega)^-$.
\end{thm}

\pagebreak

\begin{pf}
By inspection of the rules of $\TFk$, we see the following two facts.
\begin{enumerate}
 \item If a variable of large kind occurs in a derivable judgement, then a variable of large kind occurs in the context of that judgement.
\item If a variable of large kind occurs in the context of a judgement at some point in a derivation, then a variable of large kind occurs in the context of every judgement below that point.
\end{enumerate}
Therefore, if the conclusion contains no variable with large kind, then no variable with large kind occurs anywhere in the derivation, and the derivation is valid in $\mathbf{SPar}(\omega)^-$.
\end{pf}

There is a close correspondence between Canonical LF and $\mathbf{SPar}(\omega)^-$.
\begin{table}
\begin{tabular}{c c}
 Canonical LF & $\mathbf{SPar}(\omega)^-$ \\
\hline
Kinds & Product kinds of the form $(\Delta) \Type$ \\
Canonical Type Families & Product kinds of the form $(\Delta) \El{A}$ \\
Atomic Type Families & Objects of kind $\Type$ \\
Canonical Terms & Abstractions of small kind \\
Atomic Terms & Objects of small kind
\end{tabular}
\caption{Correspondence between the syntactic categories of Canonical LF and $\mathbf{SPar}(\omega)^-$.}
\label{table:clfsp}
\end{table}
It is possible to define a bijective translation between Canonical LF and $\mathbf{SPar}(\omega)^-$ that maps each class of entity in the left-hand column of Table \ref{table:clfsp} to the corresponding class of entity in the right-hand column.

The embedding of TF in LF given in this paper can be adapted in a straightforward way to provide an embedding of Canonical LF in ELF.  This embedding proves that the two systems are equivalent; that is, the derivable judgements of Canonical LF are exactly the derivable judgements of ELF that are in $\beta$-normal, $\eta$-long form.  To the best of the author's knowledge, a proof of this fact has not yet been published.  For further details, we refer to Adams \cite{adams:thesis}, where an explicit embedding of $\mathbf{SPar}(\omega)^-$ in ELF is defined.

\subsection{DMBEL}

Plotkin has produced several `algebraic frameworks' for logics and type theories, including DMBEL (Dependent Multi-Sorted Binding Equational \linebreak Logic) \cite{plotkin:afltt, pollack:srlf}.  This is a framework that allows the declaration of theories involving second-order constants, and equations between objects.  It is intended to be used for studying the theory of the syntax and semantics of logic and programming languages.  The framework DMBEL uses operations of \emph{first-order substitution} and \emph{second-order substitution}, which are similar to TF's operation of instantiation $\{M/x\}N$ restricted to the cases where $x$ is of order 0 or 1 respectively.

The framework DMBEL is essentially the same as the subsystem of $\TFk$ obtained by imposing the following restriction:
\begin{itemize}
 \item In every constant declaration, equation declaration and judgement, every variable that appears must have a small kind of order 0 or 1.
\end{itemize}
It follows that every constant that is declared must have order at most 2.  This subsystem was named $\mathbf{SPar}(2)$ in Adams \cite{adams:thesis}.  It can be proven that $\TFk$ is conservative over this subsystem:
\begin{thm}
 Let $\mathcal{T}$ be a specification in $\mathbf{SPar}(2)$, and let $\mathcal{J}$ be a judgement in which every variable has a small kind of order 0 or 1.  Then any derivation of $\mathcal{J}$ under $\mathcal{T}$ in $\TFk$ is a derivation of $\mathcal{J}$ under $\mathcal{T}$ in $\mathbf{SPar}(2)$.
\end{thm}

\begin{pf}
By inspection of the rules of $\TFk$, we see the following four facts.
\begin{enumerate}
 \item If a variable of large kind occurs in a derivable judgement, then a variable of large kind occurs in the context of that judgement.
\item If a variable of order $>1$ occurs in a derivable judgement, then a variable of order $>1$ occurs in the context of that judgement.
\item If a variable of large kind occurs in the context of a judgement at some point in a derivation, then a variable of large kind occurs in the context of every judgement below that point.
\item If a variable of order $>1$ occurs in the context of a judgement at some point in a derivation, then a variable of order $>1$ occurs in the context of every judgement below that point.
\end{enumerate}
Therefore, if the conclusion contains no variable with large kind, and no variable of order $>1$, then no variable with large kind or of order $>1$ occurs anywhere in the derivation, and the derivation is valid in $\mathbf{SPar}(2)$.
\end{pf}

There is a close correspondence between DMBEL and $\mathbf{SPar}(2)$.
\begin{table}
\begin{tabular}{c c}
DMBEL & $\mathbf{SPar}(2)$ \\
\hline
Type constant & constant of kind $(\Delta)\Type$, \\
& where $\Delta$ is small and of order $\leq 2$ \\
Term constant & constant of kind $(\Delta)\El{A}$, \\
& where $\Delta$ is small and of order $\leq 2$  \\
Term variable & variable of small kind and order 0 \\
Abstraction variable & variable of small kind and order $\leq 1$ \\
Type & object of kind $\Type$ \\
Abstraction type & small product kind of order $\leq 1$ \\
Term & object of kind $\El{A}$ \\
Abstraction term & abstraction of small kind and order $\leq 1$ \\
Context & context of order $\leq 1$ \\
Abstraction context & context of order $\leq 2$ \\
Signature & constant declarations in a specification
\end{tabular}
\caption{Correspondence between the syntactic categories of DMBEL and $\mathbf{SPar}(2)$.}
\label{table:dmbelsp}
\end{table}
It is possible to define a bijective translation between DMBEL and $\mathbf{SPar}(2)$ that maps each class of entity in the left-hand column of Table \ref{table:dmbelsp} to the corresponding class of entity in the right-hand column.

The results in this paper thus show that the properties Cut, Functionality, Equation Validity and Context Conversion hold for DMBEL, and that DMBEL can be conservatively embedded in LF.  Further, if we remove equation declarations from DMBEL, then the resulting system can be conservatively embedded in both Canonical LF and ELF.

\subsection{PAL$^+$}

The phrase `lambda-free logical framework' was originally coined to describe the framework PAL$^+$ \cite{luo:palplus}.  This framework does not use lambda-abstraction, instead taking parametrisation and local definition as primitive.  PAL$^+$ does not allow partial application; an $n$-ary function must be applied to all $n$ arguments at once.  It does still have a mechanism for forming abstractions, however; the object
\[ \operatorname{let} v[x_1 : K_1] = k : K \operatorname{in} v \]
in PAL$^+$ behaves very similarly to the lambda-abstraction $[x_1 : K_1]k$.  The system TF thus involves even fewer primitive concepts than PAL$^+$.

It can be proved that TF can be embedded in PAL$^+$, in a similar manner to the embedding in LF.  We refer to Adams \cite{adams:thesis} for the details.

\section{Conclusion}

We have presented the formal definition of two lambda-free logical frameworks, TF and $\TFk$, and proven several of their metatheoretic properties.  We have defined translations between these two frameworks and the framework LF, and shown how these can be used to lift results proven in TF to LF.

The idea of a lambda-free framework has now been invented independently by several researchers, including Aczel (who invented TF), Harper and Pfening (Canonical LF) and Plotkin (DMBEL).  These frameworks are powerful in many ways.  They represent object theories more faithfully than do traditional frameworks; each expression in the object theory corresponds to a unique object in the framework, rather than a $\beta \eta$-convertibility class.  Many results, such as the injectivity of type constructors or strong normalisation, are often easier to prove using a lambda-free framework than a traditional framework.

The cost is that the metatheoretic properties of a lambda-free framework are much more difficult to establish.  This should be seen as a one-time cost, however; these properties need only be established for a framework once, and the framework can then be used for many object theories and the lifting of many results.  We have been able to establish these properties for two large classes of object theories: those with no equation declarations, and those with only declarations of order $\leq 2$.  It follows that these results hold for Canonical LF and DMBEL, as these systems are isomorphic to conservative subsystems of TF, one of which does not allow equation declarations, and one of which does not allow specifications of order $> 2$.

For the future, the most immediate need is to remove this restriction on the specifications.  We would dearly love to be able to prove that every orderable specification is good, as we would then be able to remove the hypotheses about the $n$-goodness of specifications and the order of contexts in each of the results in this paper.
Further work should also include constructing new lambda-free logical frameworks with features such as subtyping, coercive subtyping, or meta-logical reasoning, so that results can be lifted to traditional frameworks that have these features.

\section*{Acknowledgements}

Thanks to Zhaohui Luo for helpful comments and proofreading.  Many thanks to Randy Pollack for bringing the systems CLF and DMBEL to my attention.

\bibliography{type}

\appendix

\section{Metatheory of $\mathrm{TF}$}
\label{appendix:A}
We present here the proof of the basic metatheoretic properties of TF and $\TFk$.  The proofs for each system are very similar; we shall work in TF for most of this section, and mark with the symbol \S~ the changes that need to be made to obtain a proof for $\TFk$.  These changes are all very minor.  The most substantial is in Lemma \ref{lm:bigconv1}.

Fix a natural number $n$, and let $\mathcal{T}$ be a type theory specification in TF (\S~ or $\TFk$) that is $n$-good.  Throughout this section, we assume that every kind, context, variable, constant and abstraction that appears is of order $\leq n$.

We shall begin by proving the following two properties:
\paragraph{Cut}
We say that the property Cut holds for a kind $K$ if and only if,
whenever $\Gamma, x : K, \Delta \vdash \mathcal{J}$, and $\Gamma \Vdash F : K$, then $\Gamma, \{F/x\} \Delta \vdash \{F/x\} \mathcal{J}$.
\paragraph{Functionality}
We say that the property Functionality holds for a kind $K$ if and only if,
whenever $\Gamma, x : K, \Delta \vdash M : T$ and $\Gamma \Vdash F = G : K$, then $\Gamma, \{F/x\} \Delta \vdash \{F/x\} M = \{G/x\} M : \{F/x\} T$.

We first note:
\begin{lm}
\label{lm:funceq}
Let $K$ be a kind.  Suppose the properties Cut and Functionality hold for $K$.  Then so does the following: if $\Gamma, x : K, \Delta \vdash M = N : T$ and $\Gamma \Vdash F = G : K$, then $\Gamma, \{F/x\} \Delta \vdash \{F/x\} M = \{G/x\} N : \{F/x\}T$.
\end{lm}

\begin{pf}
 Suppose $\Gamma, x : K, \Delta \vdash M = N : T$ and $\Gamma \Vdash F = G : K$.  Since the specification is good, we have $\Gamma \Vdash F : K$, and so Functionality gives
\[ \Gamma, \{ F / x \} \Delta \vdash \{ F / x \} M = \{ F / x \} N : \{ F / x \} T \enspace . \]
The goodness of the specification also gives us $\Gamma \vdash N : T$, and so 
\[ \Gamma, \{ F / x \} \Delta \vdash \{ F / x \} N = \{ G / x \} N : \{ F / x \} T \enspace . \]
The result follows by (trans).
\end{pf}

\begin{thm}
\label{thm:cutfunc}
The properties Cut and Functionality hold for every kind $K$.
\end{thm}

\begin{pf}
 The proof is by double induction, first on the kind $K$, second on the derivation of the judgement $\Gamma, x : K, \Delta \vdash \mathcal{J}$ or $\Gamma, x : K, \Delta \vdash M : T$.

\paragraph{Cut}
Let $K \equiv (\Theta)T$ and $F \equiv [\dom \Theta]P$.  (\S In $\TFk$, $F$ will have the form $[\Theta']P$.)
We deal here with the case where the last step in the derivation is
\[ (\mbox{vareq}) \; \begin{prooftree}
\Gamma, x : (\Theta) T, \Delta \Vdash \vec{H_1} = \vec{H_2} :: \Theta
 \justifies
\Gamma, x : (\Theta) T, \Delta \vdash x \vec{H_1} = x \vec{H_2} : \{ \vec{H_1} / \Theta \} T
\end{prooftree}
\]

By the induction hypothesis, we have
\[ \Gamma, \{F/x\} \Delta \Vdash \{F/x\} \vec{H_1} = \{F/x\} \vec{H_2} :: \Theta \enspace . \]
We are also given that $\Gamma, \Theta \vdash P : T$.  
By Weakening and (ref),
\[ \Gamma, \{F/x\} \Delta, \Theta \vdash P = P : T \enspace . \]
By repeatedly applying Lemma \ref{lm:funceq} with each of the kinds in $\Theta$, we have the desired conclusion
\[ \Gamma, \{F/x\} \Delta \vdash \{ \{ F / x\} \vec{H_1} / \Theta \} P = \{ \{ F / x \} \vec{H_2} / \Theta \} P : \{ \{ F / x \} \vec{H_1} / \Theta \} T \enspace . \]

\paragraph{Functionality}
Let $K \equiv (\Theta)T$, $F \equiv [\dom \Theta]P$ and $G \equiv [\dom \Theta]Q$. (\S In $\TFk$, $F$ will have the form $[\Theta']P$ and $G$ the form $[\Theta'']Q$.)  We deal here with the case where the last step in the derivation is
\[ (\mbox{var}) \; \begin{prooftree}
    \Gamma, x : (\Theta) T, \Delta \Vdash \vec{H} :: \Theta
\justifies
\Gamma, x : (\Theta) T, \Delta \vdash x \vec{H} : \{ \vec{H} / \Theta \} T
   \end{prooftree} \]

By the induction hypothesis, we have
\[ \Gamma, \{ F / x \} \Delta \Vdash \{ F / x \} \vec{H} = \{ G / x \} \vec{H} :: \Theta \enspace . \]
We are also given that
\[ \Gamma, \Theta \vdash P = Q : T \enspace . \]
By repeatedly applying Lemma \ref{lm:funceq} with each of the kinds in $\Theta$, we have the desired conclusion
\[ \Gamma, \{ F / x \} \Delta \vdash \{ \{ F / x \} \vec{H} / \Theta \} P = \{ \{ G / x \} \vec{H} / \Theta \} Q : \{ \{ F / x \} \vec{H} / \Theta \} T \enspace . \]

We also deal with the case where the last step in the derivation is
\[ (\mbox{conv}) \; \begin{prooftree}
                     \Gamma, x : K, \Delta \vdash M : \El{A}
\quad
\Gamma, x : K, \Delta \vdash A = B : \Type
\justifies
\Gamma, x : K, \Delta \vdash M : \El{B}
                    \end{prooftree} \]
We are given $\Gamma \Vdash F = G : K$; by the goodness of the specification, we also have $\Gamma \Vdash F : K$.  By the induction hypothesis, we may apply Functionality to the first premise and Cut to the second to give
\begin{gather*}
 \Gamma, \{ F / x \} \Delta \vdash \{ F / x \} M = \{ G / x \} M : \El{\{F/x\}A} \\
 \Gamma, \{ F / x \} \Delta \vdash \{ F / x \} A = \{ F / x \} B : \Type \enspace .
\end{gather*}
The result follows by (conveq).
\end{pf}

\pagebreak

Our next objective is to prove the following property:
\paragraph{Context Conversion}
We say that the property Context Conversion holds for a kind $K$ if and only if,
whenever $\Gamma, x : K, \Delta \vdash \mathcal{J}$ and $\Gamma \Vdash K = K'$, then $\Gamma, x : K', \Delta \vdash \mathcal{J}$.

Once again, we need some auxiliary lemmas:
\begin{lm}
\label{lm:Ksym}
Let $K$ be a kind, and suppose Context Conversion holds for every kind of smaller arity than $K$.
If $o(\Gamma) \leq n$ and $\Gamma \Vdash K = K'$ then $\Gamma \Vdash K' = K$.
\end{lm}

\begin{pf}
The proof is by induction on $K$.

If $K \equiv \Type$, there is nothing to prove.  If $K$ has the form $\El{A}$, we simply apply (sym).

Suppose $K \equiv (x : K_1) K_2$ and $K' \equiv (x : K_1') K_2'$.  We are given
\[ \Gamma \Vdash K_1 = K_1', \qquad \Gamma, x : K_1 \Vdash K_2 = K_2' \enspace . \]
Applying Context Conversion gives $\Gamma, x : K_1' \Vdash K_2 = K_2'$.  The desired judgements
\[ \Gamma \Vdash K_1' = K_1, \qquad \Gamma, x : K_1' \Vdash K_2' = K_2 \]
follow by the induction hypothesis.
\end{pf}

\begin{lm}
\label{lm:Csym}
 Let $\Delta$ be a context, and suppose Context Conversion holds for every kind of smaller arity than $\Delta$.  If $\Gamma \Vdash \Delta = \Delta'$ then $\Gamma \Vdash \Delta' = \Delta$.
\end{lm}

\begin{pf}
 The proof is by induction on the length of $\Delta$.  The case of length 0 is trivial.

For the inductive step, let $\Delta \equiv \Delta_0, x : K$ and $\Delta' \equiv \Delta'_0, x : K'$.  We are given
\[ \Gamma \Vdash \Delta_0 = \Delta_0' \qquad \Gamma, \Delta_0 \Vdash K = K' \enspace . \]
By the induction hypothesis, $\Gamma \Vdash \Delta_0' = \Delta_0$.  Applying Context Conversion with each of the kinds in $\Delta_0$ gives $\Gamma, \Delta_0' \Vdash K = K'$, and so $\Gamma, \Delta_0' \Vdash K' = K$ by the previous lemma.
\end{pf}

\begin{lm}
 Suppose Context Conversion holds for every kind of lower arity than $K_1$.  If $\Gamma \Vdash K_1 = K_2$ and $\Gamma \Vdash K_2 = K_3$ then $\Gamma \Vdash K_1 = K_3$.
\end{lm}

\begin{pf}
 The proof is by induction on $K_1$.  The case $K_1 \equiv \Type$ is trivial.  If $K_1$ has the form $\El{A}$, we simply apply (trans).

Suppose $K_1 \equiv (x : J_1) L_1$, $K_2 \equiv (x : J_2) L_2$, and $K_3 \equiv (x : J_3) L_3$.  We are given
\begin{xalignat*}{2}
 \Gamma & \Vdash J_1 = J_2 & \Gamma & \Vdash J_2 = J_3 \\
\Gamma, x : J_1 & \Vdash L_1 = L_2 & \Gamma, x : J_2 & \Vdash L_2 = L_3
\end{xalignat*}
By Lemma \ref{lm:Ksym}, we have $\Gamma \Vdash J_2 = J_1$; applying Context Conversion gives $\Gamma, x : J_1 \Vdash L_2 = L_3$.  The desired judgements $\Gamma \Vdash J_1 = J_3$ and $\Gamma, x : J_1 \Vdash L_1 = L_3$ follow by the induction hypothesis.
\end{pf}

\begin{lm}
\label{lm:Ctrans}
Suppose Context Conversion holds for every kind of lower arity than $\Delta_1$.  If $\Gamma \Vdash \Delta_1 = \Delta_2$ and $\Gamma \Vdash \Delta_2 = \Delta_3$, then $\Gamma \Vdash \Delta_1 = \Delta_3$.
\end{lm}

\begin{pf}
The proof is by induction on the length of $\Delta_1$.  The case of length 0 is trivial.

For the inductive step, let $\Delta_1 \equiv \Theta_1, x : K_1$; $\Delta_2 \equiv \Theta_2, x : K_2$; and $\Delta_3 \equiv \Theta_3, x : K_3$.  We are given
\begin{xalignat*}{2}
 \Gamma & \Vdash \Theta_1 = \Theta_2 & \Gamma & \Vdash \Theta_2 = \Theta_3 \\
\Gamma, \Theta_1 & \Vdash K_1 = K_2 & \Gamma, \Theta_2 & \Vdash K_2 = K_3
\end{xalignat*}
By Lemma \ref{lm:Csym}, we have $\Gamma \Vdash \Theta_2 = \Theta_1$.  Repeatedly applying Context Conversion gives us $\Gamma, \Theta_1 \Vdash K_2 = K_3$.  The desired judgements
\[ \Gamma \Vdash \Theta_1 = \Theta_3, \qquad \Gamma, \Theta_1 \Vdash K_1 = K_3 \]
follow by the induction hypothesis.
\end{pf}

\begin{lm}
\label{lm:bigconv1}
 Suppose Context Conversion holds for every kind of lower arity than $K$.  If $\Gamma \Vdash F : K$ and $\Gamma \Vdash K = K'$, then $\Gamma \Vdash F : K'$.
\end{lm}

\begin{pf}
 Let $K \equiv (\Theta) T$, $K' \equiv (\Theta') T'$ and $F \equiv [\dom \Theta]M$.  We are given
\[ \Gamma, \Theta \vdash M : T, \qquad \Gamma \Vdash \Theta = \Theta', \qquad \Gamma, \Theta \Vdash T = T' \enspace . \]
By (conv), we have $\Gamma, \Theta \vdash M : T'$.  Applying Context Conversion with each of the kinds in $\Theta$ yields
\[ \Gamma, \Theta' \vdash M : T' \]
as required.

\S In $\TFk$, let $F \equiv [\Theta_1]M$.  In addition to the above, we are given $\Gamma \Vdash \Theta = \Theta_1$ and must prove $\Gamma \Vdash \Theta' = \Theta_1$.  This follows from Lemmas \ref{lm:Csym} and \ref{lm:Ctrans}.
\end{pf}

\begin{lm}
\label{lm:bigconv}
 Suppose Context Conversion holds for each of the kinds in $\Delta$.  If $\Gamma \Vdash \vec{F} :: \Delta$ and $\Gamma \Vdash \Delta = \Delta'$, then $\Gamma \Vdash \vec{F} :: \Delta'$.
\end{lm}

\begin{pf}
 The proof is by induction on the length of $\Delta$ and $\Delta'$.  The case of length 0 is trivial.

For the induction step, let $\Delta \equiv \Delta_0, x : K$, let $\Delta' \equiv \Delta'_0, x : K'$, and let $\vec{F} \equiv \vec{F}_0, F_1$.  Then we are given
\begin{xalignat*}{2}
 \Gamma & \Vdash \vec{F}_0 :: \Delta_0 & \Gamma & \Vdash F_1 : \{ \vec{F}_0 / \Delta_0 \} K \\
\Gamma & \Vdash \Delta_0 = \Delta'_0 & \Gamma, \Delta_0 & \Vdash K = K'
\end{xalignat*}
By the induction hypothesis,
\[ \Gamma \Vdash \vec{F}_0 :: \Delta_0' \enspace . \]
Applying Cut repeatedly gives
\[ \Gamma \Vdash \{ \vec{F}_0 / \Delta_0 \} K = \{ \vec{F}_0 / \Delta_0 \} K' \]
and the desired judgement
\[ \Gamma \Vdash F_1 : \{ \vec{F}_0 / \Delta_0 \} K' \]
follows by the previous lemma.
\end{pf}

\begin{thm}
 The property Context Conversion holds for every kind $K$.
\end{thm}

\begin{pf}
Let $K \equiv (\Theta)T$ and $K' \equiv (\Theta') T'$, so we are given $\Gamma \Vdash \Theta = \Theta'$ and $\Gamma, \Theta \Vdash K = K'$.
 The proof is by double induction, first on the kind $K$, second on the derivation of $\Gamma, x : K, \Delta \vdash \mathcal{J}$.

We deal here with the case where the last step in the derivation is
\[ (\mbox{var}) \; \begin{prooftree} \Gamma, x : (\Theta) T, \Delta \Vdash \vec{F} :: \Theta
 \justifies
\Gamma, x : (\Theta) T, \Delta \Vdash x \vec{F} : \{ \vec{F} / \Theta \} T
\end{prooftree} \]

By the induction hypothesis, we have
\[ \Gamma, x : (\Theta') T', \Delta \Vdash \vec{F} :: \Theta \enspace . \]
Applying Lemma \ref{lm:bigconv}, we have
\begin{eqnarray*}
 \Gamma, x : (\Theta') T', \Delta & \Vdash & \vec{F} :: \Theta' \\
\therefore \Gamma, x : (\Theta') T', \Delta & \Vdash & x \vec{F} : \{ \vec{F} / \Theta \} T' & (var)
\addtocounter{equation}{1}\end{eqnarray*}
Applying Cut yields
\[ \Gamma, x : (\Theta') T', \Delta \vdash \{ \vec{F} / \Theta \} T = \{ \vec{F} / \Theta \} T' \]
and the result follows by (sym) and (conv).

The case where the last step is (vareq) is similar, and the other cases are all straightforward.
\end{pf}

This completes the proof of Theorem \ref{thm:mainresult}.

\paragraph{Note}
The assumption of $n$-goodness is essential for this proof.  To remove the need for it, one suggestion would be to add the following as primitive rules of TF:
\[ (\mathrm{Leq}) \; \begin{prooftree}
    \Gamma \vdash M = N : T
\justifies
\Gamma \vdash M : T
   \end{prooftree}
\qquad
(\mathrm{Req}) \;
\begin{prooftree}
 \Gamma \vdash M = N : T
\justifies
\Gamma \vdash N : T
\end{prooftree} \]
This would not work, however.  The proof of Theorem \ref{thm:cutfunc} would then fail, as we would not be able to complete the inductive step for the proof of Functionality in the case that the last step in the derivation is the rule (Req).

\section{2-good Specifications}
\label{appendix:2good}

Our aim in this section is to show that, if $\mathcal{T}$ is an orderable type theory specification in which every declaration is of order $\leq 2$, then $\mathcal{T}$ is 2-good.  In order to prove this, we must prove four properties hold simultaneously.  The following proof holds whether we are working in TF or $\TFk$.

\begin{thm}
\label{thm:2good}
Suppose $\mathcal{T}$ is an orderable specification, and
every declaration in $\mathcal{T}$ has order $\leq 2$.  Then:
\begin{enumerate}
\item Whenever $\Gamma \vdash M = N : T$ and $\Gamma$ has order $\leq 2$ then $\Gamma \vdash M : T$ and $\Gamma \vdash N : T$.
 \item Whenever $\Gamma, x : K, \Delta \vdash \mathcal{J}$, $\Gamma \Vdash F : K$ and $\Gamma, x : K, \Delta$ is of order $\leq 2$, then $\Gamma, \{ F / x \} \Delta \vdash \{ F / x \} \mathcal{J}$.
\item Whenever $\Gamma, x : K, \Delta \vdash M : T$, $\Gamma \Vdash F = G : K$ and $\Gamma, x : K, \Delta$ is of order $\leq 2$, then $\Gamma, \{ F / x \} \Delta \vdash \{ F / x \} M = \{ G / x \} M : \{ F / x \} T$.
\item Whenever $\Gamma, x : K, \Delta \vdash \mathcal{J}$, $\Gamma \Vdash K = K'$ and $\Gamma, x : K, \Delta$ is of order $\leq 2$, then $\Gamma, x : K', \Delta \vdash \mathcal{J}$.
\end{enumerate}
\end{thm}

\begin{pf}
By the orderability of $\mathcal{T}$, we may replace the rules (const), (const\_eq) and (eq) with the following rules without changing the set of derivable judgements.  For each constant declaration $c : (\Delta) T$,
\[ (\mathrm{const}') \; \begin{prooftree}
                         \Gamma \Vdash \vec{F} :: \Delta
\quad
\Delta \Vdash T \kind
\justifies
\Gamma \vdash c \vec{F} : \{ \vec{F} / \Delta \} T
                        \end{prooftree}
 \qquad (\mathrm{const\_eq}') \; \begin{prooftree}
                                  \Gamma \Vdash \vec{F} = \vec{G} :: \Delta
\quad
\Delta \vdash T \kind
\justifies
\Gamma \vdash c \vec{F} = c \vec{G} : \{ \vec{F} / \Delta \} T
\end{prooftree} \]
For each equation declaration $(\Delta)(M = N : T)$,
\[ (\mathrm{eq}') \; \begin{prooftree}
                      \Gamma \Vdash \vec{F} :: \Delta
\quad
\Delta \vdash M : T
\quad
\Delta \vdash N : T
\justifies
\Gamma \vdash \{ \vec{F} / \Delta \} M = \{ \vec{F} / \Delta \} N : \{ \vec{F} / \Delta \} T
                     \end{prooftree} \]

Given a finite sequence of declarations $s$, let us write $\Gamma \vdash_s \mathcal{J}$ to mean that there exists a derivation of the judgement $\Gamma \vdash \mathcal{J}$ such that, for every branch in the derivation, the declarations used at the (const), (const\_eq) and (eq) nodes, taken in order from leaf to root, form a subsequence of $s$.
For defined judgement forms, we write (e.g.) $\Gamma \Vdash_s (x : \El{A}) \El{B} = (x : \El{A'}) \El{B'}$ to mean $\Gamma \vdash_s A = A' : \Type$ and $\Gamma, x : A \vdash_s B = B'$ both hold.

We write $s \sqsubset t$ to denote that $s$ is a proper initial segment of $t$.  We write $\Gamma \vdash_{\sqsubset s} \mathcal{J}$ to denote that there exists $t \sqsubset s$ such that $\Gamma \vdash_t \mathcal{J}$.

Define the \emph{order} of a sequence $s$ by
\[ o(s) = \max\{o(\delta) \mid \delta \in s \} \enspace . \]

\pagebreak

We define the following properties for natural numbers $m$, $n$ with $n < m$ and sequences $s$.
\begin{itemize}
 \item $\CUT{m}{n}{s}$ is the statement: whenever $\Gamma, x : K, \Delta$ has order $m$, $K$ has order $n$, and $\Gamma, x : K, \Delta \vdash_s \mathcal{J}$ and $\Gamma \Vdash_s F : K$, then $\Gamma, \{ F / x \} \Delta \vdash_s \{ F / x \} \mathcal{J}$.
\item $\FUNC{m}{n}{s}$ is the statement: whenever $\Gamma, x : K, \Delta$ has order $m$, $K$ has order $n$, and $\Gamma, x : K, \Delta \vdash_s M : T$ and $\Gamma \Vdash_s F = G : K$, then $\Gamma, \{ F / x \} \Delta \vdash_s \{ F / x \} M = \{ G / x \} M : \{ F / x \} T$.
\item $\CC{m}{n}{s}$ is the statement: whenever $\Gamma, x : K, \Delta$ has order $m$, $K$ has order $n$, and $\Gamma, x : K, \Delta \vdash_s \mathcal{J}$ and $\Gamma \Vdash_s K = K'$, then $\Gamma, x : K', \Delta \vdash_s \mathcal{J}$.
\item $\EQVAL{m}{s}$ is the statement: whenever $\Gamma$ has order $m$ and $\Gamma \vdash_s M = N : T$, then $\Gamma \vdash_s M : T$ and $\Gamma \vdash_s N : T$. ($\mathrm{EQVAL}$ stands for `equation validity'.)
\item $\FUNCEQ{m}{n}{s}$ is the statement: whenever $\Gamma, x : K, \Delta$ has order $m$, $K$ has order $n$, and $\Gamma, x : K, \Delta \vdash_s M = N : T$ and $\Gamma \Vdash_s F = G : K$, then $\Gamma, \{ F / x \} \Delta \vdash_s \{ F / x \} M = \{ G / x \} N : \{ F / x \} T$.
\item $\GFUNC{m}{n}{s}$ is the statement: whenever $\Gamma, x : K, \Delta$ has order $m$, $K$ has order $n$, and $\Gamma, x : K, \Delta \vdash_s M : T$, $\Gamma \Vdash_s F = G : K$, $\Gamma \Vdash_s F : K$ and $\Gamma \Vdash_s G : K$, then $\Gamma, \{ F / x \} \Delta \vdash_s \{ F / x \} M = \{ G / x \} M : \{ F / x \} T$.  ($\mathrm{GFUNC}$ stands for `guarded functionality'.)
\end{itemize}
We shall employ the following abbreviations: $\CUT{\leq a}{< b}{s}$, for example, shall mean that $\CUT{m}{n}{s}$ holds for all $m \leq a$ and all $n < b$.  Another example: $\CC{m}{n}{\lessdot s}$ shall mean $\CUT{m}{n}{t}$ holds for all $t \lessdot s$.

Our aim is to show $\EQVAL{2}{s}$ for all sequences $s$ of declarations from $\mathcal{T}$.

By proofs similar to the ones in the Appendix \ref{appendix:A}, we can prove the following results for all $m$ and $s$:
\begin{enumerate}[(1)]
\item
\label{one}
$\FUNCEQ{m}{<n}{s} \wedge \CUT{m}{<n}{s} \Rightarrow \GFUNC{m}{n}{s}$
\item \label{two}
$\CUT{m}{<n}{s} \wedge \FUNCEQ{m}{<n}{s} \Rightarrow \CUT{m}{n}{s}$
\item \label{three}
$\CUT{m}{<n}{s} \wedge \CC{m}{< n-1}{s} \wedge \EQVAL{m}{s} \Rightarrow \CC{m}{n}{s}$
\item \label{four}
$\GFUNC{m}{n}{s} \wedge \CUT{m}{n}{s} \Rightarrow \GFUNCEQ{m}{n}{s}$
\end{enumerate}
The following results are trivial:
\begin{enumerate}[(1)]
\setcounter{enumi}{4}
\item \label{five}
$\GFUNC{m}{n}{s} \wedge \EQVAL{m}{s} \Rightarrow \FUNC{m}{n}{s}$
\item \label{six}
$\GFUNCEQ{m}{n}{s} \wedge \EQVAL{m}{s} \Rightarrow \FUNCEQ{m}{n}{s}$
\end{enumerate}

\pagebreak

\paragraph{Claim}
\begin{enumerate}[(1)]
\setcounter{enumi}{6}
\item 
\label{seven}
 The properties 
\begin{gather*}
\GFUNCEQ{m}{< m-1}{s} \\
\CC{m}{< m-2}{s} \\
\GFUNCEQ{\leq \max(m,o(s))}{ < o(s)}{ \sqsubset s} \\
\CC{\leq \max(m,o(s)-1)}{ < o(s)-1}{ \sqsubset s} \\
\CUT{\leq \max(m,o(s))}{<o(s)}{\sqsubset s}
\end{gather*} entail $\EQVAL{m}{s}$.
\end{enumerate}

\paragraph{Proof}
We prove that, whenever $\Gamma$ has order $\leq m$ and $\Gamma \vdash_s M = N : T$, then $\Gamma \vdash_s M : T$ and $\Gamma \vdash_s N : T$, by induction on the derivation of $\Gamma \vdash_s M = N : T$.

Suppose the last step in the derivation is
\[ (\mathrm{const\_eq}') \; \begin{prooftree}
                 \Gamma \Vdash_s \vec{F} = \vec{G} :: \Delta
\quad
\Delta \Vdash T \kind
\justifies
\Gamma \vdash_s c \vec{F} = c \vec{G} : \{ \vec{F} / \Delta \} T
                \end{prooftree} \]
where we have $(c : (\Delta)T) \in s$.  Let $s = s_1, c : (\Delta) T, s_2$, where $c : (\Delta) T$ does not occur in $s_2$.

The induction hypothesis gives $\Gamma \Vdash_{s_1} \vec{F} :: \Delta$, and so $\Gamma \vdash_s c \vec{F} : \{ \vec{F} / \Delta \} T$ by (const).

The induction hypothesis also gives $\Gamma \vdash_{s_1} G_i : \{ \vec{F} / \Delta \} K_i$, where $K_i$ is the $i$th kind in $\Delta$.  By Context Validity, we also have
\[ x_1 : K_1, \ldots, x_{i-1} : K_{i-1} \Vdash_{s_1} K_i \kind \enspace . \]
Using $\GFUNCEQ{m}{<o(s)}{\sqsubset s}$, we have $\Gamma \Vdash_{s_1} \{ \vec{F} / \Delta \} K_i = \{ \vec{G} / \Delta \} K_i$, and so, using $\CC{m}{<o(s)-1}{\sqsubset s}$,
\[ \Gamma \vdash_{s_1} G_i : \{ \vec{G} / \Delta \} K_i \enspace , \]
that is, $\Gamma \Vdash_{s_1} \vec{G} :: \Delta$.  Therefore, $\Gamma \vdash_s c \vec{G} : \{ \vec{G} / \Delta \} T$ by (const).

The case (vareq) is similar, using $\GFUNCEQ{m}{<m-1}{s}$ and \linebreak $\CC{m}{<m-2}{s}$.

Suppose $s = s_1, (\Delta)(M = N : T), s_2$, and the last step in the derivation is
\[ (\mathrm{eq}') \; \begin{prooftree}
            \Gamma \Vdash_{s_1} \vec{F} :: \Delta
\quad
\Delta \vdash_{s_1} M : T
\quad
\Delta \vdash_{s_1} N : T
\justifies
\Gamma \vdash_s \{ \vec{F} / \Delta \} M = \{ \vec{F} / \Delta \} N : \{ \vec{F} / \Delta \} T
           \end{prooftree} \]
By $\CUT{\leq \max(m,o(s))}{<o(s)}{\sqsubset s}$, we have $\Gamma \vdash_{s_1} \{ \vec{F} / \Delta \} M : \{ \vec{F} / \Delta \} T$ and $\Gamma \vdash_{s_1} \{ \vec{F} / \Delta \} N : \{ \vec{F} / \Delta \} T$.

\pagebreak

We can now use these seven results to prove Theorem \ref{thm:2good}.  Firstly, note that (1) and (2) imply
\[ \GFUNC{m}{0}{s} \wedge \CUT{m}{0}{s} \]
for every $m$ and $s$.  Therefore, by (4), $\GFUNCEQ{m}{0}{s}$ holds for every $m$ and $s$.

Our goal is to prove the following:
\[ \EQVAL{2}{s} \wedge \FUNC{2}{1}{s} \wedge \CUT{2}{1}{s} \wedge \CC{2}{1}{s} \enspace . \]
The proof is by induction on the length of $s$.  Suppose, as induction hypothesis,
\[ \EQVAL{2}{\sqsubset s} \wedge \FUNC{2}{1}{\sqsubset s} \wedge \CUT{2}{1}{\sqsubset s} \wedge \CC{2}{1}{\sqsubset s} \enspace . \]
Then the following hold:
\[ \begin{array}{cl}
\CC{2}{\leq 1}{\sqsubset s} & (\mbox{by } (\ref{three})) \\
 \EQVAL{2}{s} & (\mbox{by } (\ref{seven})) \\
\FUNC{2}{0}{s} & (\mbox{by } (\ref{five}) )\\
\FUNCEQ{2}{0}{s} & (\mbox{by } (\ref{six}) )\\
\GFUNC{2}{1}{s} & (\mbox{by } (\ref{one}) )\\
\FUNC{2}{1}{s} & (\mbox{by } (\ref{five}) )\\
\CUT{2}{1}{s} & (\mbox{by } (\ref{two}) )\\
\GFUNCEQ{2}{1}{s} & (\mbox{by } (\ref{four})) \\
\FUNCEQ{2}{1}{s} & (\mbox{by } (\ref{six}))
\end{array} \]
This completes the induction.
\end{pf}

It does not seem possible to use the same method to prove that, if every declaration in $\mathcal{T}$ is of order $\leq 3$, then $\mathcal{T}$ is 3-good.
As noted in the proof, we have $\GFUNC{m}{0}{s}$, $\CUT{m}{0}{s}$ and $\GFUNCEQ{m}{0}{s}$.  It is also possible to prove directly, by an induction on derivations, that $\CC{m}{0}{s}$ holds for all $m$ and $s$.  We are then stuck: for $o(s) = 2$, we have the circle of implications
\begin{gather*} \EQVAL{3}{s} \Rightarrow \FUNCEQ{3}{0}{s} \Rightarrow \GFUNC{3}{1}{s} \wedge \CUT{3}{1}{s} \\ \quad \Rightarrow \GFUNCEQ{3}{1}{s}
\Rightarrow \EQVAL{3}{s} \end{gather*}
without any immediate way to prove any of these directly.

We are thus unable to prove the following statement yet, and present it here as a conjecture:
\begin{conj}
Every orderable type theory specification is good.
\end{conj}

\section{The Strong Normalisability of \LF}
\label{appendix:lfsn}

Consider the simply typed lambda-calculus (STLC), with the following grammar:
\begin{eqnarray*}
 \mbox{Type}~ A & ::= & * \mid A \rightarrow A \\
\mbox{Term}~ M & ::= & x \mid \lambda x:A.M \mid MM
\addtocounter{equation}{1}\end{eqnarray*}
We shall use the fact that every term typable in STLC is strongly $\beta \eta$-normalising to prove that every object typable in LF is strongly $\beta \eta$-normalising.

Define a translation $\boxes{\,}$ that maps every kind of LF to a type of STLC, every object of LF to a term of STLC, and every context of LF to a context of STLC, as follows:
\begin{eqnarray*}
 \boxes{\Type} & \equiv & * \\
\boxes{\El{k}} & \equiv & * \\
\boxes{(x:K)K'} & \equiv & \boxes{K} \rightarrow \boxes{K'} \\
\boxes{[x:K]k} & \equiv & \lambda x : \boxes{K}. \boxes{k} \\
\boxes{kk'} & \equiv & \boxes{k} \boxes{k'} \\
\boxes{x_1 : K_1, \ldots, x_n : K_n} & \equiv & x_1 : \boxes{K_1}, \ldots, x_n : \boxes{K_n}
\addtocounter{equation}{1}\end{eqnarray*}
The key step in this proof is to realise the following fact about this translation:
\begin{lm}
 Under an arbitrary type theory specification in LF, if $\Gamma \vdash K = K'$, then $\boxes{K} \equiv \boxes{K'}$.
\end{lm}

\begin{pf}
 The proof is a simple induction on derivations.
\end{pf}

Using this lemma, we can establish the following:
\begin{lm}
\label{lm1}
 Suppose $\Gamma \vdash k : K$.  Let $c_1$, \ldots, $c_m$ be the constants that occur in $k$, and let them be declared with kinds
\[ c_1 : K_1, \ldots, c_m : K_m \enspace . \]
Then
\[ c_1 : \boxes{K_1}, \ldots, c_m : \boxes{K_m}, \boxes{\Gamma} \vdash \boxes{k} : \boxes{K} \]
in STLC.
\end{lm}

\begin{pf}
 The proof is by induction on the derivation of $\Gamma \vdash k : K$.
\end{pf}

\begin{lm}
\label{lm2}
 If $k$ and $k'$ are LF-objects and $k \rightarrow_{\beta \eta} k'$, then $\boxes{k} \rightarrow_{\beta \eta} \boxes{k'}$.
\end{lm}

\begin{pf}
We first establish the fact that
\[ \boxes{[k/x]k'} \equiv [\boxes{k} / x] \boxes{k'} \]
by induction on $k'$.  Now,
if $k \equiv ([x:K]k_1)k_2$ and $k' \equiv [k_2/x]k_1$, then
\[ \boxes{k} \equiv (\lambda x : \boxes{K}. \boxes{k_1}) \boxes{k_2} \rightarrow_\beta [\boxes{k_2} / x] \boxes{k_1} \equiv \boxes{k'} \enspace . \]
The other cases are similar.
\end{pf}

\pagebreak

These allow us to prove the theorem we want:
\begin{thm}
 Under an arbitrary type theory specification in LF, if $\Gamma \vdash k : K$, then $k$ is strongly $\beta \eta$-normalising.
\end{thm}

\begin{pf}
Suppose $k \rightarrow_{\beta \eta} k_1 \rightarrow_{\beta \eta} k_2 \rightarrow_{\beta \eta} \cdots$ is an infinite reduction sequence.
By Lemma \ref{lm1}, we have that $\boxes{k}$ is typable in STLC under some context; and by Lemma \ref{lm2}, we have that
\[ \boxes{k} \rightarrow_{\beta \eta} \boxes{k_1} \rightarrow_{\beta \eta} \boxes{k_2} \rightarrow_{\beta \eta} \cdots \]
is an infinite reduction sequence.  This contradicts the fact that STLC is strongly normalising.
\end{pf}

\end{document}